\documentclass[aps,prd,reprint,amsmath,amssymb,superscriptaddress,nofootinbib,eqsecnum,noeprint]{revtex4-2}

\usepackage{newtxtext}
\usepackage[varg]{newtxmath}
\usepackage[T1]{fontenc}
\usepackage{graphicx}
\usepackage{adjustbox}
\usepackage{soul}

\usepackage[stretch=20,shrink=10,final]{microtype}
\let\revtitle\maketitle
\renewcommand{\maketitle}{%
	\revtitle
	\tolerance=500
	\hyphenpenalty=1000
}

\renewcommand{\dot}[1]{\overset{\boldsymbol{.}}{#1}\vphantom{#1}}

\DeclareFontFamily{U}{futm}{}
\DeclareFontShape{U}{futm}{m}{n}{<-> fourier-bb}{}
\DeclareMathAlphabet{\mathbb}{U}{futm}{m}{n}

\DeclareSymbolFont{cmreg}{OT1}{cmr}{m}{n}
\DeclareSymbolFont{cmmath}{OML}{cmm}{m}{i}
\DeclareSymbolFont{cmsymbols}{OMS}{cmsy}{m}{n}
\DeclareSymbolFont{cmlargesymbols}{OMX}{cmex}{m}{n}
\DeclareSymbolFontAlphabet{\mathcal}{cmsymbols}

\DeclareMathSymbol{\partial}{0}{cmmath}{64}
\DeclareMathSymbol{g}{\mathalpha}{cmmath}{103}
\DeclareMathSymbol{\eta}{0}{cmmath}{17}
\DeclareMathSymbol{\kappa}{0}{cmmath}{20}
\DeclareMathSymbol{\mu}{0}{cmmath}{22}
\DeclareMathSymbol{\nu}{0}{cmmath}{23}
\DeclareMathSymbol{\rho}{0}{cmmath}{26}
\DeclareMathSymbol{\sigma}{0}{cmmath}{27}
\DeclareMathSymbol{\ell}{0}{cmmath}{96}

\DeclareMathSymbol{\ointop}{\mathop}{cmlargesymbols}{72}
\DeclareMathSymbol{\intop}{\mathop}{cmlargesymbols}{82}
\DeclareMathDelimiter{(}{\mathopen}{cmreg}{40}{cmlargesymbols}{0}
\DeclareMathDelimiter{)}{\mathclose}{cmreg}{41}{cmlargesymbols}{1}

\DeclareMathDelimiter{(}{\mathopen}{cmreg}{40}{cmlargesymbols}{0}
\DeclareMathDelimiter{)}{\mathclose}{cmreg}{41}{cmlargesymbols}{1}
\DeclareMathDelimiter{[}{\mathopen}{cmreg}{91}{cmlargesymbols}{2}
\DeclareMathDelimiter{]}{\mathclose}{cmreg}{93}{cmlargesymbols}{3}

\usepackage{titlesec}

\newcommand{\mysectionnumbering}{\thesection.~}

\titleformat{\section}{\bfseries\center\uppercase}{\mysectionnumbering}{0em}{}
\titleformat{\subsection}{\bfseries\center}{}{0.1em}{\thesubsection.~}
\titleformat{\subsubsection}{\bfseries\itshape\center}{}{0.1em}{\thesubsubsection.~}

\titlespacing{\section}{0pt}{1.7em plus 0.9em minus 0.9em}{0.72em plus 0.3em minus 0.2em}
\titlespacing{\subsection}{0pt}{1.5em plus 0.1em minus 0.1em}{0.5em}
\titlespacing{\subsubsection}{0pt}{1.5em plus 0.1em minus 0.1em}{0.5em}

\titleformat{\paragraph}[runin]{\itshape}{}{0em}{}[.---\,]
\titlespacing{\paragraph}{\the\parindent}{0em}{0em}

\let\oldappendix\appendix

\makeatletter
\renewcommand{\appendix}{%
\@ifstar
\oneappendix% If starred
\manyappendices% If not starred
}
\makeatother

\newcommand{\oneappendix}{%
\oldappendix*
\renewcommand{\mysectionnumbering}{\MakeUppercase{Appendix}:~}
\renewcommand\theequation{\Alph{section}\arabic{equation}}
}

\makeatletter
\newcommand{\manyappendices}{%
\oldappendix
\renewcommand{\mysectionnumbering}{\MakeUppercase{Appendix}~\thesection:~}
\renewcommand\theequation{\Alph{section}\arabic{equation}}
\renewcommand{\p@subsection}{\thesection}
}
\makeatother

\makeatletter
\renewcommand\@makefntext[1]{%
\noindent{\hspace{1em}}{\@makefnmark}#1}
\makeatother

\makeatletter
\renewcommand\@makefnmark{\hbox{\color{black}\@textsuperscript{\normalfont\@thefnmark}}}
\makeatother

\renewcommand{\footnoterule}{%
  \kern -3pt
  \hrule width 1.2cm
  \kern 4pt
}

\newcommand{\horizontalrule}[1]{\noindent\hfil\rule{#1}{0.5pt}}

\makeatletter
\def\l@@dottedsections#1#2#3#4{%
	\begingroup
	\everypar{}%
	\set@tocdim@pagenum\@tempboxa{#4}%
	\global\@tempdima\csname tocdim@#2\endcsname
	\leftskip\csname tocleft@#2\endcsname\relax
	\dimen@\csname tocleft@#1\endcsname\relax
	\parindent-\leftskip\advance\parindent\dimen@
	\rightskip\tocleft@pagenum plus 1fil\relax
	\skip@\parfillskip\parfillskip\z@
	\let\numberline\numberline@@sections
	\@nameuse{l@f@#2}%
	\ignorespaces#3\unskip\nobreak
	\leaders \hbox {$\m@th \mkern \@dotsep mu\hbox {.}\mkern \@dotsep mu$}
	\hskip\skip@
	\hb@xt@\rightskip{\hfil\unhbox\@tempboxa}\hskip-\rightskip\hskip\z@skip
	\expandafter\par
	\expandafter\aftergroup\csname tocdim@#2%
	\expandafter\endcsname
	\expandafter\endgroup
	\the\@tempdima\relax
}
\def\l@subsection{\l@@dottedsections{section}{subsection}}
\makeatother

\let\revtoc\tableofcontents
\makeatletter
\renewcommand{\tableofcontents}{%
	\onecolumngrid%
	\vskip -7pt%
	\horizontalrule{\textwidth}%
	\vskip 10pt%
	\twocolumngrid%
	\revtoc
	\onecolumngrid%
	\vskip 10pt%
	\horizontalrule{\textwidth}%
	\vskip 20pt plus 5pt%
	\twocolumngrid%
}
\makeatletter

\usepackage{etoolbox}
\apptocmd{\maketitle}{\tableofcontents}{}{}

\usepackage{xcolor}
\definecolor{revblue}{HTML}{2d3092}
\colorlet{blue}{revblue} 

\let\revcite\cite
\renewcommand\cite[1]{\mbox{\color{blue}\revcite{#1}}}

\let\reveqref\eqref
\renewcommand\eqref[1]{{\color{blue}\reveqref{#1}}}

\IfFileExists{apsrev-custom.bst}{\bibliographystyle{apsrev-custom}}{}

\newlength\mycitespacing
\setcitestyle{citesep={,\kern-\mycitespacing}}

\usepackage[
	hyperfootnotes=false,
	colorlinks=true,
	allcolors=blue,
	breaklinks=true]
{hyperref}

\usepackage{xspace}
\newcommand{\cm}{c.m.\xspace}
\newcommand{\dx}{\mathrm{d}}
\newcommand{\J}{\mathcal{J}}
\newcommand{\rad}{\text{(rad)}}
\newcommand{\stat}{\text{(stat)}}
\newcommand{\dC}{\hat{C}}
\newcommand{\iC}{\mathbb{C}}
\newcommand{\iN}{\mathbb{N}}
\newcommand{\scri}{\mathfrak{I}}
\newcommand{\id}[1]{\text{\sffamily #1}}

\newcommand{\rv}{\hat{r}}
\newcommand{\hrv}{\hat{\tilde{r}}}
\newcommand{\brv}{\rho}

\begin{document}

\title{Angular momentum balance in gravitational two-body scattering:%
\protect\\Flux, memory, and supertranslation invariance}

\author{Massimiliano Maria \surname{Riva}}
\affiliation{Deutsches Elektronen-Synchrotron DESY, Notkestr.~85, 22607 Hamburg, Germany}

\author{Filippo \surname{Vernizzi}}
\author{Leong Khim \surname{Wong}}
\affiliation{Universit\'{e} Paris-Saclay, CNRS, CEA, Institut de Physique Th\'{e}orique, 91191 Gif-sur-Yvette, France}

\begin{abstract}
Two puzzles continue to plague our understanding of angular momentum balance in the context of gravitational two-body scattering. First, because the standard definition of the Bondi angular momentum $J$ is subject to a supertranslation ambiguity, it has been shown that when the corresponding flux $F_J$ is expanded in powers of Newton's constant $G$, it can start at either $O(G^2)$ or $O(G^3)$ depending on the choice of frame. This naturally raises the question as to whether the $O(G^2)$ part of the flux is physically meaningful. The second puzzle concerns a set of new methods for computing the flux that were recently developed using quantum field theory. Somewhat surprisingly, it was found that they generally do not agree with the standard formula for $F_J$, except in the binary's center-of-mass frame. In this paper, we show that the resolution to both of these puzzles lies in the careful interpretation of $J$: Generically, the Bondi angular momentum $J$ is \emph{not} equal to the mechanical angular momentum $\mathcal{J}$ of the binary, which is the actual quantity of interest. Rather, it is the sum of $\mathcal{J}$ and an extra piece involving the shear of the gravitational field. By separating these contributions, we obtain a new balance law, accurate to all orders in $G$, that equates the total loss in mechanical angular momentum $\Delta_\mathcal{J}$ to the sum of a radiative term, which always starts at $O(G^3)$, and a static term, which always starts at $O(G^2)$. We show that each of these terms is invariant under supertranslations, and we find that $\Delta_\mathcal{J}$ matches the result from quantum field theory at least up to $O(G^2)$ in all Bondi frames. The connection between our results and other proposals for supertranslation-invariant definitions of the angular momentum is also discussed.
\end{abstract}

\maketitle

%%%%%%%%%%%%%%%%%%%%%%%%%%%%%%%%%%%%%%%%%%%%%%%%%%%%%%%%%%%%%%

\section{Introduction}
The study of gravitational two-body scattering has attracted fervent interest in recent years.~It is promising both as a theoretical arena, in which new insights into the mathematical structure of gravity may be gleaned, and as a practical tool, with which increasingly precise models of gravitational-wave signals may be developed (see, e.g., Refs.~\cite{Buonanno:2022pgc, Goldberger:2022ebt} for an~overview).
\looseness=-1

When it comes to practical calculations, the problem is generally rendered tractable by way of the post-Minkowskian expansion, which---when used alongside a number of powerful techniques adopted from high-energy physics---allows us to solve for each quantity of interest perturbatively in powers of Newton's constant~$G$~\cite{Damour:2020tta, Herrmann:2021lqe, Herrmann:2021tct, DiVecchia:2021bdo, Bjerrum-Bohr:2021din, Riva:2021vnj, Brandhuber:2021eyq, Mougiakakos:2022sic, Jakobsen:2022psy, Heissenberg:2022tsn, Riva:2022fru, Jakobsen:2022zsx, Kalin:2022hph, Dlapa:2022lmu, Bern:2022jvn, Bern:2022kto, Manohar:2022dea, DiVecchia:2022owy, DiVecchia:2022piu}. One finds in doing so that gravity manifests as a purely conservative force at first order in the approximation, and that the emission of gravitational waves appears only once we go to higher orders in~$G$.~This is easy to understand: Any Feynman diagram that contributes to the amplitude for on-shell graviton emission must contain at least one internal graviton line, each of which confers a factor of~$G$, and one external graviton~leg, which adds an additional factor of~$G^{1/2}$~\cite{Jakobsen:2021smu, Mougiakakos:2021ckm}. Since the total flux of four-momentum~$F_P$ is proportional to the square of this amplitude~\cite{Goldberger:2016iau}, $F_P$~must start at~$O(G^3)$.~This simple power-counting argument is corroborated by explicit calculations of the four-momentum flux, which were first carried out for the case of two point masses in Refs.~\cite{Herrmann:2021lqe, Herrmann:2021tct, DiVecchia:2021bdo, Bjerrum-Bohr:2021din, Riva:2021vnj}, before being generalized to include tidal interactions~\cite{Mougiakakos:2022sic, Jakobsen:2022psy, Heissenberg:2022tsn}~and~spin~effects~\cite{Riva:2022fru, Jakobsen:2022zsx}.

Explicit calculations \cite{Herrmann:2021tct, DiVecchia:2021bdo, Kalin:2022hph, Jakobsen:2022psy, Jakobsen:2022zsx} also confirm that the total flux of four-momentum radiated across future null infinity is precisely equal to the total change in the four-momenta of the two bodies.
This kind of \emph{balance law}, in which changes in the mechanical properties of the binary are linked to the flux of outgoing radiation, provides an important consistency check and, in the case of bound orbits, plays a key role in the construction of waveform models~\cite{Damour:2011fu, Blanchet:2013haa, Blanchet:2018yqa, Ashtekar:2019viz}.

One expects a similar balance law to hold for the angular momentum, but it is here that we encounter two puzzles. First, applying the same power-counting argument from before naively predicts that the angular momentum flux~$F_J$ should also start at $O(G^3)$, but explicit calculations~\cite{Damour:2020tta, Manohar:2022dea, DiVecchia:2022owy, DiVecchia:2022piu} reveal that it actually begins one order earlier, at~$O(G^2)$. It was understood in Ref.~\cite{Damour:2020tta} that this $O(G^2)$ part of the flux is linked to the gravitational-wave memory, and can be interpreted as saying that angular momentum is also transferred, starting at $O(G^2)$, into the static components of the gravitational field. (In the language of particle physics, one says that it is transferred into zero-frequency gravitons~\cite{DiVecchia:2022owy, DiVecchia:2022piu}.) Meanwhile, the transfer of angular momentum into radiative modes remains an~$O(G^3)$~effect, as explained in Ref.~\cite{Veneziano:2022zwh}. This~state of affairs is not entirely intuitive, but it is also not necessarily a problem.~Indeed, something similar happens in electromagnetism~\cite{Bern:2021xze, Cristofoli:2022phh}, for which the relevant expansion parameter is the fine-structure constant. The real puzzle arises when we confront this result with the coordinate freedom that general relativity~affords.

Because the Bondi angular momentum~$J$ is subject to a supertranslation ambiguity~\cite{doi:10.1007/s10714-010-1110-5, Newman:1966ub, doi:10.1063/1.524242, Ashtekar:2019rpv}, it turns out that a suitable change of coordinates (amounting to a pure supertranslation) can be used to remove the $O(G^2)$ part of the flux entirely~\cite{Veneziano:2022zwh}. This then raises the question as to whether the $O(G^2)$ part of the flux is physically meaningful, or if it is merely a coordinate artifact. Two findings support the notion that it is physical. The first is a linear-response relation between the conservative and radiation-reaction parts of the scattering angle~\cite{Bini:2012ji, Damour:2020tta} (see~also Ref.~\cite{DiVecchia:2021ndb, Jakobsen:2022zsx}), which produces the correct result at~$O(G^3)$ only if the angular momentum flux starts at~$O(G^2)$. The second is a set of explicit solutions to the binary's equations of motion~\cite{Damour:1981bh, Bini:2022wrq}, which assert that the binary always loses mechanical angular momentum starting at~$O(G^2)$. Taken together, these various results present us with a puzzle of why a seemingly physical part of the flux can be set to zero by a change of coordinates, and why the loss of mechanical angular momentum from the binary may or may not be balanced by the angular momentum flux, depending on the choice of coordinate frame. (See also Ref.~\cite{Veneziano:2022zwh} for further~discussion.)

The second puzzle concerns the explicit computation of this angular momentum flux.~Owing to a number of recent advancements \cite{Manohar:2022dea, DiVecchia:2022owy, DiVecchia:2022piu}, there are now at least two different approaches that one could take. The first is to use a classic formula, given by Thorne~\cite{Thorne:1980ru, DeWitt:2011nnj}, that yields the space-space components of the flux~$\smash{ F^{ij}_J }$ after an integration over position space. This approach was adopted in Refs.~\cite{Damour:2020tta, Jakobsen:2021smu, Mougiakakos:2021ckm} to compute~$\smash{ F^{ij}_J }$ up to~$O(G^2)$.~The second approach involves a set of new formulas \cite{Manohar:2022dea, DiVecchia:2022owy, DiVecchia:2022piu}, based on quantum field theory, that yield both the space-space and time-space components of the flux after an integration over momentum space. Results obtained via this second approach are available up to~$O(G^3)$~\cite{Manohar:2022dea, DiVecchia:2022owy, DiVecchia:2022piu}.~Surprisingly, where a comparison is possible, these two approaches generally do not agree~\cite{Manohar:2022dea}, except in the binary's center-of-mass~(\cm) frame. It has been suggested that a possible explanation for this discrepancy is the inapplicability of Thorne's formula outside the \cm~frame, but this contradicts the fact that Thorne's formula can be obtained from the Bondi-Sachs formalism \cite{Bondi:1960jsa, Bondi:1962px, Sachs:1962wk, Sachs:1962zza} without imposing any restrictions on the binary's~\cm~motion~\cite{Compere:2019gft}.

In this paper, we will instead argue that the resolution to both puzzles lies in the careful interpretation~of~$J$. After reviewing several key aspects of the Bondi-Sachs formalism in Sec.~\ref{sec:bondi}, we show in Sec.~\ref{sec:mech} that the Bondi angular momentum~$J$ of a system is generically \emph{not} equal to its mechanical angular momentum~$\J$, which is the quantity we actually care about. Case in point, consider a single Schwarzschild black hole of mass $m$ whose center of energy travels along the worldline ${x^\mu(\tau) = b^\mu + p^\mu \tau/m}$. The constant vector~$b^\mu$ is the displacement of this worldline from the spacetime origin, $p^\mu$ is its four-momentum, and $\tau$ is the proper time. It is then the mechanical angular momentum~$\J$ that is given by the familiar formula ${ \J^{\mu\nu} = 2 b^{[\mu} p^{\nu]} }$. The Bondi angular momentum~$J$, on the other hand, can be written as the sum of $\J$ and an extra term that depends on the shear of the gravitational field. 

For the two-body case, we use this general relation between $J$ and~$\J$ to derive a new balance~law, accurate to all orders in~$G$, that equates the total loss of mechanical angular momentum from the binary to the sum of a radiative term and a static term.~If $\smash{ \J_{- \mathstrut}^{\mu\nu} }$ and $\smash{ \J_{+ \mathstrut}^{\mu\nu} }$ denote the values of the mechanical angular momentum before and after the scattering event, respectively,~then%
\begin{gather}
	\J^{\mu\nu}_{+ \mathstrut} - \J^{\mu\nu}_{- \mathstrut}
	=
	-\Delta_\J^{\mu\nu},
	\nonumber\\
	\Delta_\J^{\mu\nu}
	\equiv
	\Delta_{\J\rad}^{\mu\nu}
	+
	\Delta_{\J\stat}^{\mu\nu}.
\label{eq:balance_mech_J_Lorentz}
\end{gather}
The radiative term~$\smash{ \Delta_\J^\rad }$ [Eq.~\eqref{eq:loss_mech_J_rad}], which always starts at~$O(G^3)$, accounts for the flux of angular momentum that is carried away by gravitational waves, while $\smash{ \Delta_\J^\stat }$ [Eq.~\eqref{eq:loss_mech_J_stat}], which always starts at $O(G^2)$, accounts for the additional transfer of angular momentum into the static components of the gravitational field. The physical significance of the former has previously been appreciated in Refs.~\cite{Veneziano:2022zwh, Javadinezhad:2022ldc}, but identifying how the latter arises from the Bondi-Sachs formalism is a key contribution of this work. Both of these terms are inherently physical, as we show that $\J$, $\smash{\Delta_\J^\rad}$, and~$\smash{\Delta_\J^\stat}$ are all individually invariant under pure supertranslations. On the whole, we consider these results to be a satisfactory resolution to the first of~our~two~puzzles.

A resolution to the second puzzle is provided in Sec.~\ref{sec:pm}. After computing the total loss $\Delta_\J$ for a two-body scattering event explicitly at~$O(G^2)$, we verify that it agrees with the result from quantum field theory~\cite{Manohar:2022dea, DiVecchia:2022owy, DiVecchia:2022piu} at this order in all Bondi frames.~Additionally, we find that $\smash{\Delta_\J^\stat}$ matches the corresponding static part of the result in Refs.~\cite{DiVecchia:2022owy, DiVecchia:2022piu} also at~$O(G^3)$. These results establish that the reason for the general discrepancy between Refs.~\cite{Manohar:2022dea, DiVecchia:2022owy, DiVecchia:2022piu} and Refs.~\cite{Jakobsen:2021smu, Mougiakakos:2021ckm} is that they are, in fact, computing two different quantities: the former references compute~$\Delta_\J$, whereas the latter compute the Bondi flux~$F_J$.~We are also able to explain why the space-space components of $F_J$ and $\Delta_\J$ just so happen to agree at $O(G^2)$ in the binary's \cm~frame. Avenues for future work are discussed alongside our conclusions~in~Sec.~\ref{sec:conclusion}.

Complementing the main text are four appendices that address some of the more technical aspects of this paper. In Appendix~\ref{app:Lorentz}, we show how to translate between the scalar-valued integrals $\{ P(\sigma), J(\sigma), \dots \}$ of the Bondi-Sachs formalism and the more familiar representation of the momenta and their fluxes as Lorentz tensors, $\smash{ \{ P^\mu_{\mathstrut}, J^{\mu\nu}_{\mathstrut}, \dots \} }$. Because scalars are considerably easier to work with, our presentation in the main text will mostly favor use of the former, although several occasions will arise when switching to the latter becomes beneficial. The more tedious steps involved in our derivation of $\J$ and $\Delta_\J$ are collected in Appendices~\ref{app:schwz} and~\ref{app:proofs}, and finally, in Appendix~\ref{app:comparisons} we compare our results with other recent proposals for a supertranslation-invariant definition of the angular~momentum~\cite{Compere:2019gft, Chen:2021szm, Chen:2021kug, Mao:2023evc, Javadinezhad:2022hhl, Javadinezhad:2022ldc}.

Our metric signature is ($-$,$+$,$+$,$+$), our antisymmetrization convention is such that ${T^{[\mu\nu]} = (T^{\mu\nu} - T^{\nu\mu})/2}$, and we adopt units in~which~${c = 1}$~throughout. 

% ---------------------------------------------------------- %
\section{Bondi-Sachs formalism}
\label{sec:bondi}

This section provides a brief introduction to the Bondi-Sachs formalism~\cite{Bondi:1960jsa, Bondi:1962px, Sachs:1962wk, Sachs:1962zza}, which is well suited to the study of radiation in asymptotically flat spacetimes. We begin by discussing the general form of the Bondi metric near future null infinity in Sec.~\ref{sec:bondi_metric}, before turning to an enumeration of its asymptotic symmetries in Sec.~\ref{sec:bondi_BMS}. The link between  asymptotic symmetries and balance laws is then explored in Sec.~\ref{sec:bondi_balance}. Our exposition is mostly an abridged version of Refs.~\cite{Madler:2016xju, Flanagan:2015pxa, Bonga:2018gzr}, to which we refer the reader for more details.

\subsection{Bondi metric}
\label{sec:bondi_metric}

When seeking to describe the transport of radiation towards future null infinity~$\scri^+$, it is convenient to choose a coordinate chart that is adapted to outgoing null rays. The retarded Bondi coordinates $(u,r,\theta^A)$, with ${ A \in \{1,2\} }$, form one such example. In these coordinates, the hypersurfaces of constant retarded time~$u$ are taken to be null, while the angular coordinates~$\theta^A$ are defined such that every null ray that is tangent to one of these hypersurfaces is a curve along which $u$, $\theta^1$, and~$\theta^2$ are constant. The remaining radial coordinate~$r$ then parametrizes our position along a given null ray.

On its own, this construction imposes only three constraints on the metric, namely ${g^{uu} = g^{u A} = 0}$~\cite{Madler:2016xju}. We remove the last remaining gauge degree of freedom by also requiring that ${\partial_r \det (g_{AB}/r^4) = 0}$, which forces the coordinate~$r$ to be the areal radius. The most general metric that we can write down subject to these constraints is~then
\begin{align}
	\dx s^2
	=&
	- \mu e^{2\beta}\dx u^2 -2e^{2\beta}\dx u\dx r
	\nonumber\\&
	+
	\gamma_{AB}
	(r\dx\theta^A + W^A \dx u)
	(r\dx\theta^B + W^B \dx u).
\label{eq:Bondi_metric}
\end{align}

Asymptotic flatness is imposed by requiring that this metric reduces to that of Minkowski in the limit~${r \to \infty}$. The appropriate boundary conditions on the metric components are thus ${\mu \to 1}$, ${\beta \to 0}$, ${W^A \to 0}$, and ${\gamma_{AB} \to \Omega_{AB}}$, where $\Omega_{AB}$ is the round metric on the unit 2-sphere; i.e., ${\Omega_{AB} = \mathrm{diag}(1,\sin^2\theta)}$ in the usual $(\theta,\phi)$~chart. For large but finite values of the radius, we can expand these metric components in powers of $1/r$ to obtain an accurate description of the spacetime in the vicinity of~$\scri^+$~\cite{Bondi:1962px, Sachs:1962wk}. Assuming for simplicity that the metric in Eq.~\eqref{eq:Bondi_metric} satisfies the vacuum Einstein equations in this region (this~does not preclude the existence of matter but merely requires that it be concentrated away from~$\scri^+$), we find that the most relevant terms in the expansion are~\cite{Flanagan:2015pxa}
\begin{subequations}
\label{eq:Bondi_metric_expansion}
\begin{align}
	\mu &= 1 - \frac{2GM}{r} + O(r^{-2}),
	\allowdisplaybreaks\\
	\gamma_{AB} &= \Omega_{AB} + \frac{1}{r} C_{AB}
	+
	O(r^{-2}),
	\allowdisplaybreaks\\
	\beta &= -\frac{1}{32r^2} C_{AB} C^{AB} + O(r^{-3}),
	\allowdisplaybreaks\\
	W^A &= \frac{1}{2r} D_B C^{AB}
	+
	\frac{1}{r^2}
	\bigg(
		\frac{2}{3} GN^A - \frac{1}{16} D^A (C^{BC} C_{BC})
		\nonumber\\&\quad
		+
		\frac{1}{2} C^{AB} D^C C_{BC}
	\bigg)
	+
	O(r^{-3}),
\end{align}
\end{subequations}
where indices are always raised and lowered with~$\Omega_{AB}$, and $D_A$ is the covariant derivative compatible with~$\Omega_{AB}$.

We see from Eq.~\eqref{eq:Bondi_metric_expansion} that the spacetime is fully characterized at this order in $1/r$ by just three objects: the~mass aspect~$M$, which has dimensions of mass; the angular momentum aspect~$N_A$,%
\footnote{Different papers use slightly different definitions for the angular momentum aspect; for a summary, see Eqs.~(2.8) and~(2.9) of Ref.~\cite{Compere:2018ylh}. Our definition coincides with that of Flanagan and Nichols~\cite{Flanagan:2015pxa}.}
which has dimensions of angular momentum; and the shear tensor~$C_{AB}$, which has dimensions of length. All three objects are functions only of the three coordinates~$(u, \theta^A)$, and we note that the shear tensor must be traceless (i.e.,~${\Omega^{AB} C_{AB} = 0}$) as a consequence of the gauge~constraint~on~$r$.

The vacuum Einstein equations also govern how two of these quantities evolve with time. Writing~$\smash{\dot X \equiv \partial_u X}$ for any quantity~$X$, the evolution equation for~$M$~reads
\begin{equation}
	G\dot M = - \frac{1}{8} N_{AB} N^{AB} + \frac{1}{4} D_A D_B N^{AB},
\label{eq:Einstein_M}
\end{equation}
while the corresponding equation for~$N_A$~is
\begin{align}
	G \dot N_A
	=&\;
	G D_A M
	+
	\frac{1}{4} D_B D_A D_C C^{BC} - \frac{1}{4} D^2D^B C_{AB}
	\nonumber\\&
	+
	\frac{1}{4} D_B(N^{BC} C_{CA}) + \frac{1}{2} C_{AB} D_C N^{BC}.
\label{eq:Einstein_N}
\end{align}
Both equations depend on the news tensor,
\begin{equation}
	N_{AB} \coloneq \dot{C}_{AB},
\end{equation}
but there is no third equation that independently constrains the evolution of~$C_{AB}$. This makes intuitive sense, because the shear tensor is where the information about gravitational waves is encoded, and we have yet to specify any details about the gravitational-wave source. For the case of two-body scattering, these details might come from, say, calculating the amplitude for on-shell graviton emission. In any case, once a solution for ${C_{AB} \equiv C_{AB}(u,\theta^A)}$ is provided, Eqs.~\eqref{eq:Einstein_M} and~\eqref{eq:Einstein_N} automatically dictate how $M$ and~$N_A$ evolve away from their~initial~conditions.

\subsection{Asymptotic symmetries}
\label{sec:bondi_BMS}

We call a coordinate chart $(u,r,\theta^A)$ a ``Bondi frame'' if it results in a metric of the general form in Eqs.~\eqref{eq:Bondi_metric} and~\eqref{eq:Bondi_metric_expansion}. The asymptotic symmetries of~$\scri^+$ may then be defined as those coordinate transformations ${(u,r,\theta^A) \mapsto (u',r',\theta'^A)}$ that take us from one Bondi frame to another. These transformations form the Bondi-Metzner-Sachs~(BMS) group~\cite{Bondi:1962px, Sachs:1962wk, Sachs:1962zza}, which is structurally similar to the Poincar\'{e} group, except that the four-dimensional subgroup of translations is replaced by an infinite-dimensional subgroup of ``supertranslations''.

This structure is readily seen by considering a general element of the Lie algebra. For an infinitesimal transformation with ${u' = u + \xi^u}$, ${r' = r + \xi^r}$, and ${\theta'^A =\theta^A + \xi^A}$, one finds that the asymptotic Killing vector~\cite{Flanagan:2015pxa}
\begin{align}
	\xi
	=&\;
	\xi^u \partial_u + \xi^r \partial_r + \xi^A \partial_A
	\allowdisplaybreaks\nonumber\\
	=&\:
	\bigg(
		\alpha + \frac{1}{2}u D_A Y^A
	\bigg)
	\partial_u
	-
	\bigg(
		\frac{1}{2} r D_A Y^A + O(r^0)
	\bigg)
	\partial_r
	\nonumber\\[0.25em]&
	+
	\big[
		Y^A + O(r^{-1})
	\big]
	\partial_A
\end{align}
is parametrized by one arbitrary scalar function~${\alpha\equiv\alpha(\theta^A)}$ and one vector ${Y^A \equiv Y^A(\theta^B)}$, which must satisfy the conformal Killing~equation%
\footnote{Although we do not consider them here, it is worth mentioning that there are several extensions of the BMS algebra~\cite{Barnich:2009se, *Barnich:2010eb, *Barnich:2011mi, Campiglia:2014yka, *Campiglia:2015yka}, which impose less stringent constraints on the vector~$Y^A$.}
\begin{equation}
	2 D_{(A}Y_{B)} - (D_C Y^C)\Omega_{AB} = 0 . 
\label{eq:conformal_Killing}
\end{equation}
The claim is that $Y^A$ parametrizes the Lorentz transformations, while~$\alpha$ parametrizes the supertranslations.

On flat spacetimes, we are typically accustomed to seeing an infinitesimal Lorentz transformation ${x'^\mu = x^\mu + \omega^\mu{}_\nu x^\nu}$ being generated by the constant antisymmetric tensor~$\omega_{\mu\nu}$, and an infinitesimal translation ${x'^\mu = x^\mu - a^\mu}$ being generated by the constant vector~$a^\mu$.~The physical significance of~$Y^A$ and~$\alpha$ can thus be made clearer if we are able to express them in terms of these more familiar objects. We do so by introducing the Lorentzian coordinates ${x^\mu \equiv (t, x, y, z)}$, whose time coordinate ${t=u+r}$ and whose spatial coordinates $(x,y,z)$ are related to the spherical Bondi coordinates $(r,\theta,\phi)$ in the usual~way. Then~defining
\begin{equation}
\label{eq:n_def}
	n^\mu \coloneq (1, \sin\theta\cos\phi, \sin\theta\sin\phi, \cos\theta)
\end{equation}
as the outgoing radial null vector on~$\scri^+$ and $\bar{n}^\mu$ as its image under the antipodal map~${(\theta,\phi) \mapsto (\pi-\theta, \phi + \pi)}$, it is possible to write the general solution to Eq.~\eqref{eq:conformal_Killing} as~\cite{Flanagan:2015pxa}
\begin{equation}
	Y_A
	= \omega_{\mu\nu} n^\mu \partial_A \bar{n}^\nu,
\label{eq:Y_Lorentz}
\end{equation}
where Greek indices are always raised and lowered with the Minkowski metric~$\eta_{\mu\nu}$.

For the supertranslations, we use the fact that any function on the \mbox{2-sphere} is decomposable into spherical harmonics to write ${\alpha = \alpha_{\ell\leq 1} + \alpha_{\ell\geq 2}}$, where $\alpha_{\ell\leq 1}$ is formed by an appropriate linear combination of the ${\ell=0}$ and ${\ell=1}$ harmonics, while $\alpha_{\ell\geq 2}$ is formed by the remaining harmonics with ${\ell\geq 2}$. That we~can~always~write
\begin{equation}
	\alpha_{\ell\leq 1} = a_\mu n^\mu
\label{eq:st_Lorentz}
\end{equation}
for some~$a_\mu$ establishes this part of~$\alpha$ as being responsible for the standard translations; the remaining part $\alpha_{\ell\geq 2}$ generates the pure supertranslations.

\subsection{Charges and fluxes}
\label{sec:bondi_balance}

Noether's theorem tells us that the four-momentum~$P^\mu$ and angular momentum~$J^{\mu\nu}$ are the ten conserved charges associated with the Poincar\'{e} symmetries of Minkowski space. For asymptotically flat spacetimes, a general prescription due to Wald and Zoupas~\cite{Wald:1999wa} provides the analog of this result by associating a charge to every BMS generator~$\xi$.%
\footnote{Alternative prescriptions for defining charges \cite{Dray:1984rfa} and fluxes \cite{Ashtekar:1981bq} are known to yield the same result~\cite{Wald:1999wa} (see also Ref.~\cite{Elhashash:2021iev}).}
The four-momentum and supermomentum charges, which are conjugate to translations~$\alpha_{\ell\leq 1}$ and pure supertranslations~$\alpha_{\ell\geq 2}$, respectively, are both encoded in the surface~integral~\cite{Flanagan:2015pxa}
\begin{equation}
	P(\sigma) = \int_\sigma\frac{\dx^2\Omega}{4\pi}\,\alpha M.
\label{eq:def_BMS_P}
\end{equation}
Meanwhile, the angular momentum charge, which is conjugate to the Lorentz transformations~$Y^A$, is given by%
\footnote{Different papers adopt slightly different conventions for the~numerical factors appearing in front of the two terms quadratic in the shear tensor, leading to a two-parameter family of definitions; see Refs.~\cite{Compere:2019gft, Elhashash:2021iev} for details. The definition in Eq.~\eqref{eq:def_BMS_J} is the unique one that (i)~vanishes on flat space~\cite{Elhashash:2021iev} and (ii)~is balanced by a corresponding flux $F_J$ whose expression matches the classic formula by Thorne~\cite{Thorne:1980ru}. It is nevertheless possible to relax condition~(ii) and still arrive at the same relation in Eq.~\eqref{eq:def_BMS_J_split} between $J$ and the mechanical angular momentum~$\J$. This is because the terms quadratic in the shear tensor vanish whenever $C_{AB}$ can be written in the form of Eq.~\eqref{eq:schwz_metric_components_CAB}~\cite{Elhashash:2021iev}. As per Eq.~\eqref{eq:twobody_CAB}, we see that the initial and final states of a binary undergoing scattering exhibit this property.}
\cite{Flanagan:2015pxa}
\begin{align}
	J(\sigma)
	=
	\int_\sigma\frac{\dx^2\Omega}{8\pi G}&\,Y^A
	\bigg(
		G\hat N_A
		-
		\frac{1}{16}D_A(C_{BC} C^{BC})
		\nonumber\\&
		-
		\frac{1}{4} C_{AB} D_C C^{BC}
	\bigg),
\label{eq:def_BMS_J}
\end{align}
where, for later convenience, we have introduced the shifted angular momentum aspect 
\begin{equation}
	\hat{N}_A \coloneq N_A - u D_A M.
\end{equation}

\begin{figure}
\centering\includegraphics{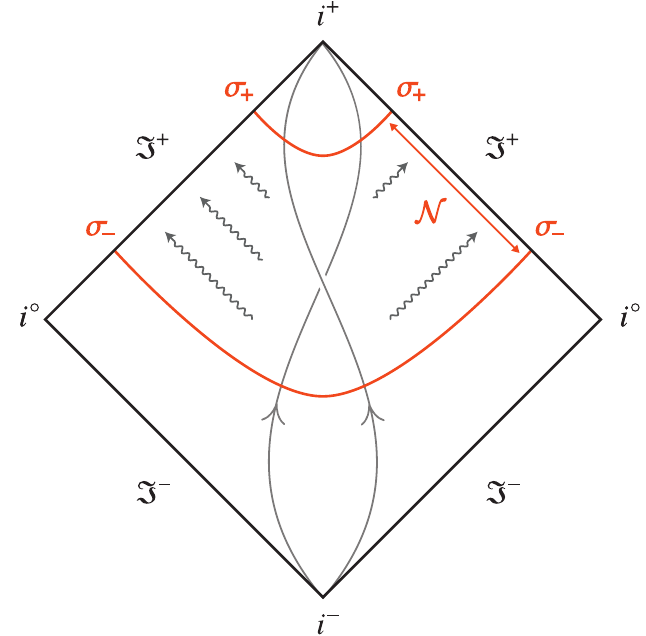}
\caption{Penrose diagram for the asymptotically flat spacetime around a binary undergoing scattering. The centers of energy of the binary's constituents trace out worldlines that travel from past timelike infinity~$i^-$ to future timelike infinity~$i^+$, while the gravitational waves that they emit travel towards future null infinity~$\scri^+$. It is assumed that no incoming radiation travels from past null~infinity~$\scri^-$. Also drawn on this diagram are spatial infinity~$i^\circ$, and two asymptotically null hypersurfaces that intersect~$\scri^+$ at the cuts~$\sigma_-$ and~$\sigma_+$. We denote the region of~$\scri^+$ bounded between $\sigma_-$ and $\sigma_+$ by~$\mathcal{N}$.}
\label{fig:Penrose}	
\end{figure}

Notice, crucially, that these charges are defined on a given ``cut''~$\sigma$, which is a \mbox{2-sphere} of constant~$u$ on~$\scri^+$, because their values generally change with time as the system emits gravitational radiation. The total change between two cuts, say $\sigma_-$ and $\sigma_+$, is determined by the balance~laws
\begin{subequations}
\label{eq:balance_laws}
\begin{gather}
	P(\sigma_+) - P(\sigma_-) = - F_P(\mathcal{N}),
	\label{eq:balance_P}
	\\
	J(\sigma_+) - J(\sigma_-) = - F_J(\mathcal{N}),
	\label{eq:balance_J}
\end{gather}
\end{subequations}
where $\mathcal{N}$ is the region of~$\scri^+$  bounded between $\sigma_-$ and~$\sigma_+$, as illustrated in Fig.~\ref{fig:Penrose}. The total flux of four-momentum and supermomentum is given by~\cite{Flanagan:2015pxa}
\begin{equation}
	F_P(\mathcal{N})
	=
	\int_\mathcal{N} \frac{ \dx u \dx^2\Omega }{32\pi G}\,\alpha
	(N^{AB} N_{AB} - 2D_A D_B N^{AB}),
\label{eq:flux_P}
\end{equation}
while the total flux of angular momentum is%
\footnote{The formula for the angular momentum flux in Eqs.~(C4) and~(C5) of Ref.~\cite{Flanagan:2015pxa} contains an extra term in the integrand of the form~$u Y^A D_A D_B D_C N^{BC}$. This term does not contribute to the flux if $Y^A$ is restricted to be part of the standard BMS algebra, as we do here. To see this, note that three successive integrations by parts can be used to rewrite this term as $-u N^{BC} D_C D_B D_A Y^A$, which vanishes after use of the identity in Eq.~\eqref{eq:id_Y_3}.}
\cite{Flanagan:2015pxa}
\begin{align}
	F_J(\mathcal{N})
	=&\;
	\int_\mathcal{N} \frac{ \dx u \dx^2\Omega }{32\pi G}\,Y^A
	\bigg(
		N^{BC} D_A C_{BC}
		-
		2D_B(N^{BC} C_{AC})
		\nonumber\\&
		+
		\frac{1}{2} D_A(N^{BC} C_{BC})
		-
		\frac{1}{2} u D_A(N^{BC} N_{BC})
	\bigg).
\label{eq:flux_J}
\end{align}

It is not difficult to verify that the balance law for $P(\sigma)$ is consistent with the Einstein equations; multiplying Eq.~\eqref{eq:Einstein_M} by $\alpha / (4\pi G)$ and then integrating over the region~$\mathcal{N}$ easily reproduces Eq.~\eqref{eq:balance_P}. The balance law for $J(\sigma)$ can also be shown to be consistent with Eqs.~\eqref{eq:Einstein_M} and~\eqref{eq:Einstein_N}, although the steps are more involved~\cite{Flanagan:2015pxa}.

To conclude this section, it is worth remarking that the parametrizations for $Y^A$ and $\alpha$ in Eqs.~\eqref{eq:Y_Lorentz} and~\eqref{eq:st_Lorentz} enable us to convert between the scalar-valued integrals $\{ P(\sigma), J(\sigma)$, $F_P(\mathcal N), F_J(\mathcal N) \}$ of the Bondi-Sachs formalism and the more familiar representation of these charges and their fluxes as Lorentz tensors. We define the four-momentum $P^\mu$ and the angular momentum~$J^{\mu\nu}$ on a given cut~$\sigma$~via
\begin{equation}
	P^\mu(\sigma) = \frac{\partial P(\sigma)}{\partial a_\mu},
	\quad
	J^{\mu\nu}(\sigma) = \frac{\partial J(\sigma)}{\partial\omega_{\mu\nu}},
\label{eq:Lorentz_charges}
\end{equation}
and for the fluxes radiated across~$\mathcal{N}$, we~define
\begin{equation}
	F_P^\mu(\mathcal{N}) = \frac{F_P(\mathcal{N})}{\partial a_\mu},
	\quad
	F_J^{\mu\nu}(\mathcal{N}) = \frac{\partial F_J(\mathcal{N})}{\partial\omega_{\mu\nu}}.
\label{eq:Lorentz_fluxes}
\end{equation}
[In the same way, the supermomentum charges and fluxes can be extracted by differentiating~$P(\sigma)$ and $F_P(\mathcal{N})$ with respect to a suitable parametrization of $\alpha_{\ell\geq 2}$~\cite{Flanagan:2015pxa}, but for our purposes these quantities do not play a role.]

In Appendix~\ref{app:Lorentz}, we show that it is also possible to map the individual metric components $\{ M, N_A, C_{AB}, N_{AB} \}$ onto a corresponding set of pseudotensors $\{ M, N_\mu, C_{\mu\nu}, N_{\mu\nu} \}$. Explicit expressions for the fluxes in terms of these objects are then given in Eqs.~\eqref{eq:flux_P_Lorentz} and~\eqref{eq:flux_J_Lorentz}. Finally, we note here that by writing ${ C_{\mu\nu} = \lim_{r \to \infty} r h_{\mu\nu}^\text{TT} }$, one can verify that the integrands of $\smash{F_P^\mu}$ and $\smash{F_J^{\mu\nu}}$ are in agreement with the differential flux formulas for the energy, linear momentum, and space-space components of the angular momentum as given~by~Thorne~\cite{Thorne:1980ru}.%
\footnote{The space-space components of the angular momentum flux are given in the form $\smash{\epsilon_{ijk} F^{jk}_J /2}$ in Eq.~($4.22'$) of Ref.~\cite{Thorne:1980ru}, where $\epsilon_{ijk}$ is the Levi-Civita symbol; no formula is given for the remaining time-space components~$\smash{ F^{0i}_J }$, which are associated with changes in the position of the system's center of mass. While a multipole-expanded version of~$\smash{F^{0i}_J}$ can be found in, e.g., Refs.~\cite{Nichols:2018qac, Blanchet:2018yqa}, the manifestly Lorentz-covariant formula for $\smash{F_J^{\mu\nu}}$ presented in Eq.~\eqref{eq:flux_J_Lorentz} appears~to~be~new.}

% ---------------------------------------------------------- %
\section{Mechanical angular momentum}
\label{sec:mech}

This section introduces our definition for the mechanical angular momentum~$\J$ and establishes some of its key properties. To motivate this definition, we begin in Sec.~\ref{sec:mech_schwz} by considering the special case of a single Schwarzschild black hole moving at constant velocity. This simple example is instructive because we know \emph{a~priori} what the value of~$\J$ should~be. Accordingly, we show that the Bondi angular momentum~$J$ for this spacetime is generically not equal to~$\J$, but contains an extra piece that depends on the shear of the gravitational field.

We then generalize this result in Sec.~\ref{sec:mech_twobody} to the problem of two-body scattering by determining analogous relations between $J$ and~$\J$ in the initial and final states. In Sec.~\ref{sec:mech_balance}, these relations are combined with Eq.~\eqref{eq:balance_J} to obtain a new balance law, accurate to all orders in~$G$, that equates the total loss of mechanical angular momentum from the binary to the sum of two terms: one describing a flux of radiation, and another describing a static effect associated with the gravitational-wave memory [see~Eq.~\eqref{eq:balance_mech_J}]. The behavior of these two terms under finite BMS transformations is then examined in Sec.~\ref{sec:mech_supertranslation}, where we find, in particular, that they are individually invariant under~supertranslations.

\subsection{Boosted black hole}
\label{sec:mech_schwz}

Consider a Schwarzschild black hole of mass $m$ moving at a constant velocity with respect to a Lorentzian coordinate chart ${\tilde{x}^{ \mu} }$. For concreteness, we assume this  chart to be harmonic,%
\footnote{Note that the results of this subsection are not unique to harmonic coordinates, however. We obtain the same end result when repeating this exercise by starting with the Schwarzschild metric in isotropic coordinates, for~instance.}
i.e.,~to satisfy the conditions ${\tilde g^{\rho \sigma} \tilde \nabla_\rho \tilde \nabla_\sigma \tilde{x}^{\mu}=0}$, where $\tilde \nabla_\mu$ is the covariant derivative compatible with the metric $\tilde g_{\mu\nu}$ that describes the boosted black hole spacetime in these coordinates. (This choice of coordinates is particularly relevant for the two-body case, since practical post-Minkowskian calculations are invariably done in harmonic or de~Donder coordinates.) To obtain this metric, we boost and translate the static Schwarzschild metric in harmonic coordinates so that the worldline of the black hole's center of energy in the $\tilde x^\mu$ chart is given by ${\tilde x^{\mu}(\tau) = b^{\mu} + p^{\mu}\tau / m}$, where $\tau$ is the black hole's proper time, $p^\mu$ its four-momentum, and $b^\mu$ the displacement of this worldline from the spacetime origin.%
\footnote{Although the harmonic coordinates do not extend past the event horizon, we can infer by extrapolation that the black hole's center of energy is located at the origin of the coordinate chart in which the Schwarzschild metric is static and spherically symmetric. Boosting and translating to the $\tilde x^\mu$ chart then implies that the black hole's worldline is given by ${\tilde x^{ \mu}(\tau) = b^{ \mu} + p^{ \mu} \tau / m}$.}
Our goal in this subsection is to construct a general definition for the mechanical angular momentum that correctly evaluates to the expected result ${\J^{\mu\nu} = 2b^{[\mu} p^{\nu]}}$ in this special~case.

To proceed, we need to know how the Lorentz vectors $b^\mu$ and $p^\mu$ enter into the components $\{ M, N_A, C_{AB} \}$ of the Bondi metric. We accomplish this by transforming the metric components in harmonic coordinates, which explicitly depend on $b^\mu$ and $p^\mu$, into the metric components in Bondi coordinates. We then match the quantities $\{ M, N_A, C_{AB} \}$ after performing an expansion in powers of~$1/r$. The full details of this calculation are given in Appendix~\ref{app:schwz}. (See also Refs.~\cite{Blanchet:2020ngx, Veneziano:2022zwh} for related  derivations.)

Crucially, because the partial differential equations that determine the coordinate transformation from the harmonic metric to the Bondi metric are all linear [see~Eqs.~\eqref{eq:NU_eqs_O1} and~\eqref{eq:NU_eqs_O2}], their general solution  must therefore be  the sum of a particular integral and a complementary function. The former is the part of the transformation that actually takes us from harmonic to Bondi gauge, while the latter corresponds to a residual gauge freedom that exists once we are already in Bondi gauge: This is precisely the freedom to perform a BMS transformation from one Bondi frame to another~\cite{Blanchet:2020ngx}. Since we do not want to change the physical state of the system by boosting ourselves into a new frame in which the black hole travels at a different velocity, we shall set the part of the complementary function associated with Lorentz transformations to zero.~It will be instructive, however, to keep the part of the complementary function associated with supertranslations arbitrary for the time being; we shall parametrize this part by the scalar function~${\beta\equiv\beta(\theta^A)}$.

Having done so, we find that the resulting Bondi metric for a boosted Schwarzschild spacetime~has%
\begin{subequations}
\label{eq:schwz_metric_components}
\begin{align}
	M &= m^4 / (- n \cdot p)^3,
	\label{eq:schwz_metric_components_M}
	\\
	\hat{N}_A
	&=
	3M D_A (B+S) + (B+S) D_A M,
	\label{eq:schwz_metric_components_N}
	\\
	C_{AB}
	&=
	-(2 D_A D_B - \Omega_{AB} D^2) S,
	\label{eq:schwz_metric_components_CAB_S}
\end{align}
where the scalar functions
\begin{align}
	B &= (n \cdot b),
	\\
	S &=
	2G (n\cdot p)
	\log\left( \frac{-n\cdot p}{m} \right) + \beta.
	\label{eq:schwz_metric_components_S}
\end{align}
\end{subequations}
The null vector $n^\mu$ is as defined in Eq.~\eqref{eq:n_def}, and inner products like ${n \cdot p \equiv \eta_{\mu\nu} n^\mu p^\nu}$ are always taken with respect to the Minkowski metric on~$\scri^+$. Different subsets of the above result can be found across Refs.~\cite{Bondi:1962px, Bonga:2018gzr, Compere:2019gft, Veneziano:2022zwh}.

It will be helpful in what follows to decompose the function $S$ into spherical harmonics. We write ${S=Z+C}$, where ${Z \equiv S_{\ell\leq 1}}$ contains only the ${\ell\leq 1}$ harmonics of~$S$, while ${C \equiv S_{\ell\geq 2}}$ contains the remaining ${\ell\geq 2}$ harmonics. This decomposition is useful because $Z$ lives in the kernel of the differential operator ${(2D_A D_B - \Omega_{AB}D^2)}$, and so Eq.~\eqref{eq:schwz_metric_components_CAB_S} may equivalently be written~as
\begin{equation}
	C_{AB}
	=
	-(2 D_A D_B - \Omega_{AB} D^2) C.
\label{eq:schwz_metric_components_CAB}
\end{equation} 
We call~$C$ the ``shear'' of the gravitational field, since it serves as a kind of potential for the shear tensor~$C_{AB}$.

The metric components in Eq.~\eqref{eq:schwz_metric_components} determine the Bondi charges $P$ and~$J$. For~the former, we insert Eq.~\eqref{eq:schwz_metric_components_M} into Eq.~\eqref{eq:def_BMS_P}, differentiate with respect to~$a_\mu$ as per Eq.~\eqref{eq:Lorentz_charges}, and then integrate over the angular coordinates (which is easily done by, e.g., choosing the spatial part of $p^\mu$ to point along the $z$~axis) to find that ${P^\mu = p^\mu}$, i.e., that the Bondi four-momentum of this spacetime is precisely equal to the mechanical four-momentum of the black hole. This is not surprising, but the point is worth laboring because the same is not true of the~angular~momentum.

To obtain the Bondi angular momentum~$J$, we first note that the terms in Eq.~\eqref{eq:def_BMS_J} that are quadratic in $C_{AB}$ cancel one another upon insertion of Eq.~\eqref{eq:schwz_metric_components_CAB}~\cite{Elhashash:2021iev}, and thus the integrand of $J$ depends only on the shifted angular momentum aspect~$\hat N_A$. Substituting in Eq.~\eqref{eq:schwz_metric_components_N}, we then find it natural to split the result into three~parts.~We~write
\begin{equation}
	J = j(M, B) + j(M, Z) + j(M, C),
\label{eq:def_BMS_J_split}
\end{equation}
where, for any two scalar functions $f_1$~and~$f_2$, we~define
\begin{equation}
	j(f_1, f_2)
	=
	\int\frac{\dx^2\Omega}{8\pi}
	Y^A (3 f_1 D_A f_2 + f_2 D_A f_1).
\label{eq:def_little_j}
\end{equation}

Switching to the Lorentz-tensor representation makes the physical significance of the first term in Eq.~\eqref{eq:def_BMS_J_split} apparent; we show in Appendix~\ref{app:Lorentz}~that
\begin{equation}
	\frac{\partial j(M,B)}{\partial\omega_{\mu\nu}}
	=
	2 b^{[\mu} p^{\nu]},
\label{eq:def_mech_J_schwz_Lorentz}	
\end{equation}
which is the desired result for what we want to call the mechanical angular momentum~$\J$.

We now turn our attention to the second term in Eq.~\eqref{eq:def_BMS_J_split}. Because~$Z$ is composed of ${\ell\leq 1}$ harmonics only, there exists a constant vector~$z^\mu$ such that ${Z = (n \cdot z)}$. This means that $j(M, B+Z)$ becomes $2(b^{[\mu} + z^{[\mu}) p ^{\nu]}$ in the Lorentz-tensor representation, and thus the function~$Z$ corresponds to an additional translation that is generated when we transform from harmonic to Bondi coordinates. From Eq.~\eqref{eq:schwz_metric_components_S}, we see that $Z$ would be a $p^\mu$-dependent translation were we to set the complementary function~$\beta$ to zero. To remove this spurious translation that is introduced by the particular integral, we learn that the appropriate boundary condition to impose is to choose $\beta_{\ell\leq 1}$ such that~${Z=0}$.

Setting ${Z=0}$ now leaves us with the relation
\begin{equation}
	J = \J + j(M,C),
\label{eq:def_BMS_J_split_Z=0}
\end{equation}
which says that the Bondi angular momentum~$J$ of this spacetime is given by the sum of its mechanical angular momentum~$\J$ and an extra term that depends on the shear of the gravitational field. Written out explicitly, we have that
\begin{equation}
 C = \mathbb{P}_{\ell\geq 2}\!
 	\left[
 		2G (n\cdot p) \log\left( \frac{-n\cdot p}{m} \right)
 	\right]
 	+
 	\beta_{\ell\geq 2},
\end{equation}
where $\mathbb{P}_{\ell\geq 2}$ is a projection operator that keeps the spherical harmonic modes with~${\ell\geq 2}$ only, and recall that we have yet to impose any restrictions on the complementary function~$\beta_{\ell\geq 2}$. Indeed, any choice of $\beta_{\ell\geq 2}$ corresponds to a valid Bondi frame, as the dependence of $j(M,C)$ on $\beta_{\ell\geq 2}$ is precisely the well-known ambiguity of~$J$ under (pure)~supertranslations.

\subsection{Two-body scattering}
\label{sec:mech_twobody}

The full spacetime for the scattering encounter between two massive bodies is undoubtedly more complicated than the single black hole spacetime considered in the previous subsection, but its limiting form in the asymptotic past and future is tame enough that we can still make quantitative statements about the binary in its initial and final state. On~the two cuts ${ \sigma_\pm \to \scri^+_\pm }$, where $\scri^+_-$ and~$\scri^+_+$ denote the past and future endpoints of future null infinity, respectively, we define the mechanical angular momentum of the binary implicitly via the relation [cf.~\eqref{eq:def_BMS_J_split}]
\begin{equation}
	J^\pm = \J^\pm + j(M^\pm, Z^\pm) + j(M^\pm, C^\pm),
\label{eq:def_mech_J_binary}
\end{equation}
where we write ${X^\pm \equiv X(\scri^+_\pm)}$ for any quantity~$X$; this relation is guaranteed by requiring that $J$ be equal to~$\J$ when evaluated in a canonical frame. Since the total loss of Bondi angular momentum ${J^- - J^+}$ is given by Eq.~\eqref{eq:balance_J}, we can determine the mechanical angular momentum loss ${\J^- - \J^+}$ once we have specified appropriate boundary values for the mass aspect $M^\pm$ and the scalar functions~${S^\pm = Z^\pm + C^\pm}$. Crucially, notice that this can be done without the need for explicit expressions of~$\J^\pm$. (An analogous discussion on the initial and final states of an inspiraling binary can be found in~Refs.~\cite{Compere:2019gft, Ashtekar:2019rpv, Ashtekar:2019viz}.)

At both $\scri^+_-$ and~$\scri^+_+$, the binary consists of two widely separated bodies traveling along asymptotically straight trajectories. We expect the gravitational binding energy between these bodies to become negligible as their spacelike separation goes to infinity~\cite{Compere:2019gft}; hence, $M^\pm$ must be given by a superposition of the individual mass aspects for the two bodies. Said in other words, if $m_a$, $p^\mu_{a-}$, and $p^\mu_{a+}$ denote the rest mass, ingoing four-momentum, and outgoing four-momentum of the $a$th body, respectively,~then
\begin{equation}
	M^\pm
	=
	\sum_{a=1}^2 \frac{m_a^4}{(- n \cdot p_a^\pm)^3}.
\label{eq:twobody_M_initial_and_final}
\end{equation}

Since we showed in Sec.~\ref{sec:mech_schwz} that the Bondi four-momentum $P^\mu$ for a single black hole is precisely its mechanical four-momentum~$p^\mu$, it follows from the form of Eq.~\eqref{eq:def_BMS_P} and our assumption of the principle of superposition~that
\begin{equation}
	P_\pm^\mu = \sum_{a=1}^2 p_{a\pm}^\mu
\end{equation}
for the case of two-body scattering. The balance law
\begin{equation}
	P_+^\mu - P_-^\mu = - F_P^\mu
\label{eq:balance_P_Lorentz}
\end{equation}
then implies that the sum of the individual losses of mechanical four-momentum from each body is exactly equal to the total flux of four-momentum radiated across future null~infinity. Explicit post-Minkowskian calculations have verified that this is the case~\cite{Herrmann:2021tct, DiVecchia:2021bdo, Kalin:2022hph, Jakobsen:2022psy, Jakobsen:2022zsx}. [Note~that we have suppressed the argument on $\smash{ F_P^\mu \equiv F_P^\mu(\scri^+) }$ in Eq.~\eqref{eq:balance_P_Lorentz}; in the rest of this paper, it is to be understood that fluxes are always being evaluated along the~entirety~of~$\scri^+$.]

Similar steps are used to determine~$S^\pm$. As we did for the mass aspect, we shall assume that we can use the principle of superposition to say that
\begin{equation}
	S^-
	=
	\sum_{a=1}^2
	2G (n\cdot p_a^-)
	\log\left( \frac{-n\cdot p_a^-}{m_a} \right)
	+
	\beta.
\label{eq:twobody_S_initial}
\end{equation}
This agrees with the result first obtained in~Ref.~\cite{Veneziano:2022zwh} for the generic $N$-body case. The first term in Eq.~\eqref{eq:twobody_S_initial} is just the particular integral that we would obtain by starting with a superposition of two boosted black holes in harmonic coordinates and then transforming to Bondi coordinates, while the second is the arbitrary complementary function~$\beta$ associated with supertranslations. As in Sec.~\ref{sec:mech_schwz}, we shall eliminate the spurious $p_a^\mu$-dependent translation generated by the particular integral by imposing appropriate boundary conditions on $\beta_{\ell\leq 1}$ such that
\begin{subequations}
\label{eq:twobody_Z_C_initial}
\begin{equation}
	Z^- \equiv S_{\ell\leq 1}^- = 0.
\label{eq:twobody_Z_initial}
\end{equation}

As for the initial shear of the gravitational field, we have that
\begin{equation}
	C^-
	=
	\sum_{a=1}^2
 	\mathbb{P}_{\ell\geq 2}\!
 	\left[
 		2G (n\cdot p_a^-) \log\left( \frac{-n\cdot p_a^-}{m_a} \right)
 	\right]
 	+
 	\beta_{\ell\geq 2},
\label{eq:twobody_C_initial}
\end{equation}
\end{subequations}
with $\beta_{\ell\geq 2}$ still arbitrary. 
In later parts of this section, we will show that the total loss of mechanical angular momentum~$\Delta_\J$ is independent of the value of~$\beta_{\ell\geq 2}$ (i.e., it is invariant under pure supertranslations), but it will nevertheless be useful to introduce two common gauge-fixing choices.  Following the language of Ref.~\cite{Veneziano:2022zwh}, we define the ``intrinsic gauge'' as the family of Bondi frames in which ${\beta_{\ell\geq 2} = 0}$,%
\footnote{In Ref.~\cite{Veneziano:2022zwh}, the ``intrinsic gauge'' is used to refer to a Bondi frame with ${\beta=0}$, including ${\beta_{\ell \leq 1}=0}$. We refer the reader to this reference for a more detailed and physical explanation of this gauge. In this work, we set ${Z^-=0}$, which is just a different choice of origin for the mechanical angular momentum~$\J^-$, and this means that ${\beta_{\ell\leq 1}\neq 0}$. Despite this difference, we adopt the same terminology as in Ref.~\cite{Veneziano:2022zwh} and call this the intrinsic gauge because the pure supertranslation ambiguity is fixed in the same way, i.e., by setting ${\beta_{\ell \geq 2}=0}$.}
and we define the ``canonical gauge'' as the family of Bondi frames in which $\beta_{\ell\geq 2}$ is chosen such that~${C^-=0}$.%
\footnote{Note that reference~\cite{Flanagan:2015pxa} uses the term ``canonical'' in a stronger sense to mean a frame in which $\smash{\dot M}$, $D_A M$, $\smash{\dot N_A}$, and $C_{AB}$ are all zero.}
As can be seen from Eq.~\eqref{eq:def_BMS_J_split_Z=0}, the latter is particularly useful because the Bondi angular momentum~$J^-$ is equal to the mechanical angular momentum~$\J^-$ when evaluated in a~canonical~frame~\cite{Veneziano:2022zwh}. However, for the sake of generality, we shall leave $\beta_{\ell\geq 2}$ unspecified to show how it drops out of the final result.

The result for~$S^+$ is more subtle. At~low orders in the post-Minkowskian expansion, $S^+$ must be identical to~$S^-$, except with $p_a^-$ replaced by~$p_a^+$, because scattering processes are symmetric under time reversal in the absence of radiation. When radiation is included, this must mean that
\begin{equation}
	S^+
	=
	\sum_{a=1}^2
	2G (n\cdot p_a^+)
	\log\left( \frac{-n\cdot p_a^+}{m_a} \right)
	+
	\beta
	+
	O(G\Delta\mathcal{E}),	
\label{eq:twobody_S_final}
\end{equation}
where we write $O(G\Delta\mathcal{E})$ to signify the presence of additional terms associated with the emission of gravitational waves (see~also~Ref.~\cite{Damour:2020tta}); the quantity $\Delta\mathcal{E}$, which is defined below Eq.~\eqref{eq:DeltaM}, is the total energy radiated per unit solid angle across~$\scri^+$. These extra terms are linked to the nonlinear part of the gravitational-wave memory, and are discussed in further detail in Sec.~\ref{sec:pm_stat}.

For now, we turn our attention back to the first two terms in Eq.~\eqref{eq:twobody_S_final}. Observe that the function~$\beta$ in this equation is the same function that appears in Eq.~\eqref{eq:twobody_S_initial}. This is no accident, because we cannot perform different supertranslations at different times---the total loss ${\J^- - \J^+}$ is a meaningful quantity only when $\J^-$ and $\J^+$ are both evaluated in the same Bondi frame. Two key implications follow from this restriction. The first is that ${Z^+ \neq 0}$ because we have already fixed the value of $\beta_{\ell\leq 1}$ to set ${Z^- = 0}$. Instead, we are left with
\begin{subequations}
\label{eq:twobody_Z_C_final}
\begin{equation}
	Z^+ =
	\sum_{a=1}^2
	\mathbb{P}_{\ell\leq 1}\!
 	\left[
 		2G (n\cdot p_a) \log\left( \frac{-n\cdot p_a}{m_a} \right)
 	\right]_{-\infty}^{+\infty}
 	+
 	O(G\Delta\mathcal{E}),
 \label{eq:twobody_Z_final}
\end{equation}
where we write ${[X]^{+\infty}_{-\infty} \equiv X^+ - X^-}$ for brevity. Physically, this result suggests that the reference point about which the Bondi angular momentum~$J$ is defined inadvertently shifts as the system emits gravitational waves.~The term $j(M^+, Z^+)$ in Eq.~\eqref{eq:def_mech_J_binary} corrects for this shift, such that the mechanical angular momenta $\J^+$ and $\J^-$ are defined with respect to the~same~origin.

The second implication is that ${C^+ \neq 0}$ even if we choose to work in the canonical gauge wherein ${C^- = 0}$. More generally, since the final value of the shear
\begin{equation}
	C^+
	=
	\sum_{a=1}^2
 	\mathbb{P}_{\ell\geq 2}\!
 	\left[
 		2G (n\cdot p_a^+) \log\left( \frac{-n\cdot p_a^+}{m_a} \right)
 	\right]
 	+
 	\beta_{\ell\geq 2}
 	+
 	O(G\Delta\mathcal{E}),
\label{eq:twobody_C_final}
\end{equation}
\end{subequations}
we see by subtracting Eq.~\eqref{eq:twobody_C_initial} from the above that the difference ${\Delta C = C^+ - C^-}$ is independent of~$\beta_{\ell\geq 2}$, and so is invariant under supertranslations. The fact that this quantity cannot be set to zero by a coordinate transformation is a hint that it is physical, and indeed it is well known that the tensor $\smash{ \Delta C_{AB} \equiv -(2 D_A D_B - \Omega_{AB} D^2) \Delta C }$ is responsible for the gravitational-wave memory effect~\cite{Zeldovich:1974gvh, doi:10.1038/327123a0, Christodoulou:1991cr, Wiseman:1991ss, Blanchet:1992br, Thorne:1992sdb, Bieri:2013ada, Tolish:2014oda, Strominger:2014pwa, Garfinkle:2022dnm}, whereby a permanent change to the relative displacement between two freely falling observers is induced by the passage of a gravitational~wave. Accordingly, in what follows we shall refer to $\Delta C$ as the ``gravitational memory.''

Two more remarks are worth making at this stage. First, because the two scalar functions ${C^\pm \equiv C^\pm(\theta^A)}$ are independent of the retarded time~$u$, the corresponding shear~tensors
\begin{equation}
	C^\pm_{AB} = -(2 D_A D_B - \Omega_{AB} D^2) C^\pm
\label{eq:twobody_CAB}
\end{equation}
are also independent of~$u$, and thus ${N_{AB} = 0}$ at~$\scri^+_\pm$. This alone is insufficient as a boundary condition on the news tensor, however. To ensure that the fluxes $F_P$ and~$F_J$ remain finite, we will have to impose the stricter requirement that ${N_{AB} \sim O(|u|^{-1-\epsilon})}$ for some ${\epsilon > 0}$~as~${|u| \to \infty}$.

The second remark pertains to the universality of the results in Eqs.~\eqref{eq:twobody_M_initial_and_final}, \eqref{eq:twobody_Z_C_initial}, and~\eqref{eq:twobody_Z_C_final}. While our construction of these asymptotic data made specific use of the Bondi metric for a boosted Schwarzschild black hole, these results are nevertheless valid for binary systems composed of any type of body---black holes, neutron stars, white dwarfs, etc.---spinning or otherwise. The reason is that the functions $M$, $Z$, and~$C$ are part of the leading-order terms in the $1/r$ expansion of the Bondi metric, and so are sensitive only to the mass monopoles of the two bodies.~Consequently, we emphasize that the new balance law in Eq.~\eqref{eq:balance_mech_J}, which we will derive shortly, makes no assumptions about the nature of the two bodies, nor do the usual balance laws in Eq.~\eqref{eq:balance_laws}.

\subsection{Balance law}
\label{sec:mech_balance}

Now,~combining Eqs.~\eqref{eq:balance_J} and~\eqref{eq:def_mech_J_binary} and then simply rearranging terms, we see that
\begin{equation}
	\J^+ - \J^- = - \Delta_\J,
\end{equation}
where we define the total mechanical angular momentum loss
\begin{equation}
	\Delta_\J
	\coloneq
	F_J + j(M^+,Z^+) + j(M^+,C^+) - j(M^-,C^-).
\label{eq:def_total_loss}
\end{equation}
What remains is to understand the meaning of this~definition.

Because the function $j(\cdot\,,\cdot)$ is bilinear in its two arguments, we may equivalently~write
\begin{equation}
	j(M^+,C^+) - j(M^-,C^-)
	=
	j(\Delta M, C^-) + j(M^+,\Delta C), 
\label{eq:j_bilinear_identity}
\end{equation}
where ${\Delta M = M^+ - M^-}$ and recall that ${\Delta C = C^+ - C^-}$. Proceeding with the term $j(\Delta M, C^-)$, we first integrate by parts to move the derivative off~$\Delta M$. Total divergences vanish since a 2-sphere has no boundary; hence,
\begin{equation}
	j(\Delta M, C^-)
	=
	\int\frac{\dx^2\Omega}{8\pi}\,
	(2 Y^A D_A C^- - C^- D_A Y^A)
	\Delta M .
\label{eq:j_deltaM_C-}
\end{equation}
We now use the Einstein equations to express $\Delta M$ as a function of the shear. Integrating Eq.~\eqref{eq:Einstein_M} with respect to~$u$~yields
\begin{equation}
	\Delta M = \frac{1}{4G} D^A D^B \Delta C_{AB} - \Delta\mathcal{E},
\label{eq:DeltaM}
\end{equation}
where $\smash{ \Delta\mathcal{E} \coloneq (1/8G)\int_{- \infty}^{+ \infty}  N^{AB} N_{AB} \dx u }$ is the total energy radiated per unit solid angle across~$\scri^+$, while $\smash{\Delta C_{AB} \equiv \int_{-\infty}^{+\infty} N_{AB} \dx u}$. By inserting this into Eq.~\eqref{eq:j_deltaM_C-}, integrating by parts, and then using a number of identities as outlined in Appendix~\ref{app:proofs}, we eventually find that
\begin{align}
	j(\Delta M, C^-)
	=&
	-\!
	\int\frac{\dx u \dx^2\Omega}{32\pi G}
	\bigg(
		N^{BC} D_A C^-_{BC}
		-
		2 D_B(N^{BC} C_{AC}^-)
		\nonumber\\&
		+
		\frac{1}{2} D_A (N^{BC} C^-_{BC})
		+
		\frac{1}{2} D_A(N^{BC} N_{BC} C^-)
		\nonumber\\&
		+
		N^{BC} N_{BC} D_A C^-
	\bigg)
	Y^A.
\label{eq:j_deltaM_Cplus_final}
\end{align}

Observe that the first three terms in the integrand are almost identical to the first three terms in~$F_J$ [see~Eq.~\eqref{eq:flux_J}], except that $C_{AB}$ is here replaced by~$\smash{ -C_{AB}^- }$. By~adding these two equations together, we are naturally led to define
\begin{align}
	\Delta_\J^\rad
	\coloneq &\;
	F_J + j(\Delta M, C^-)
	\nonumber\\
	=&\;
	\int\frac{\dx u \dx^2\Omega}{32\pi G} Y^A
	\bigg(
		N^{BC} D_A \dC_{BC}
		-
		2D_B(N^{BC} \dC_{AC})
		\nonumber\\&
		+
		\frac{1}{2} D_A(N^{BC} \dC_{BC})
		-
		\frac{1}{2} u D_A(N^{BC} N_{BC})
		\nonumber\\&
		-
		\frac{1}{2} D_A(N^{BC} N_{BC} C^-)
		-
		N^{BC} N_{BC} D_A C^-
	\bigg)
\label{eq:loss_mech_J_rad}
\end{align}
as the \emph{radiated flux} of mechanical angular momentum,~where
\begin{align}
	\dC_{AB}^{\mathstrut}(u,\theta^C)
	\coloneq &\;
	C_{AB}^{\mathstrut}(u,\theta^C) - C_{AB}^{-}(\theta^C)
	\nonumber\\
	=&
	\int_{-\infty}^u \dx u' N_{AB}(u', \theta^C)
\label{eq:def_dynamical_C}
\end{align}
denotes the dynamical part of the shear tensor~\cite{Javadinezhad:2022hhl}.

Two considerations justify our interpretation of $\smash{\Delta_\J^\rad}$ as the radiated flux. The first is that Eq.~\eqref{eq:loss_mech_J_rad} depends explicitly on the radiative modes of the gravitational field, via~$N_{AB}$ and~$\dC_{AB}$, at all intermediate times ${u \in (-\infty, +\infty)}$. The second reason is that Eq.~\eqref{eq:loss_mech_J_rad} starts at $O(G^3)$ when expanded perturbatively in powers of~$G$. To see this, we use Eq.~\eqref{eq:flux_P_Lorentz} and the fact that the four-momentum flux $\smash{ F_P^\mu }$ is known to start at~$\smash{ O(G^3) }$ to deduce that $\smash{ N_{AB} }$ must start at~$O(G^2)$. It~then follows from the second line of Eq.~\eqref{eq:def_dynamical_C}~that $\dC_{AB}$ also starts at~$O(G^2)$.~The value of the initial shear~$C^-$ is arbitrary, however, because it depends on the arbitrary function~$\beta_{\ell\geq 2}$ [see~Eq.~\eqref{eq:twobody_C_initial}], and so the last two terms in Eq.~\eqref{eq:loss_mech_J_rad} stand to ruin our power counting scheme. Fortunately, it~turns out that these two terms are exactly what is needed to render $\smash{\Delta_\J^\rad}$ invariant under supertranslations. We~show this explicitly in Sec.~\ref{sec:mech_supertranslation}, but for now, the implication is that $\smash{\Delta_\J^\rad}$ does not actually depend on~$C^-$; hence, we can set ${C^- = 0}$ without loss of generality to conclude that $\smash{\Delta_\J^\rad}$ always starts~at~$O(G^3)$.

Returning to Eqs.~\eqref{eq:def_total_loss} and~\eqref{eq:j_bilinear_identity}, we recall that $\smash{\Delta_\J^\rad}$ is not the only contribution to the mechanical angular momentum loss. We group the remaining terms into what we call the \emph{static contribution},
\begin{align}
	\Delta_\J^\stat
	&\coloneq
	j(M^+, \Delta C) + j(M^+, Z^+)
	\nonumber\\
	&=
	\int\!\frac{\dx^2\Omega}{8\pi}
	M^+
	\left(
		2 Y^A D_A \Delta S - \Delta S D_A Y^A
	\right),
\label{eq:loss_mech_J_stat}
\end{align}
where ${\Delta S \equiv \Delta C + \Delta Z}$ and ${\Delta Z \equiv Z^+}$, since ${Z^- = 0}$.~This object depends only on quantities defined on $\scri^+_\pm$, namely, the final value of the mass aspect~$M^+$, the gravitational memory~$\Delta C$, and the translation~$\Delta Z$ that corrects for the shift in reference point about which the Bondi angular momentum~$J$ is defined. To count the powers of~$G$ that appear in this expression, we first write ${M^+ = M^- + \Delta M}$. The initial mass aspect~$M^-$ is independent of~$G$ as it depends only on the ingoing four-momenta of the two bodies, while $\Delta M$ starts later at~$O(G)$, since the mechanical impulse ${p_{a+}^\mu - p_{a-}^\mu = O(G)}$ [see~Eq.~\eqref{eq:impulse_1PM}]. As for~$\Delta S$, we see from Eqs.~\eqref{eq:twobody_Z_C_initial} and~\eqref{eq:twobody_Z_C_final} that it starts at $O(G^2)$; hence, this static term~$\smash{\Delta_\J^\stat}$, which like~$\smash{\Delta_\J^\rad}$ is also invariant under supertranslations, always~starts~at~$O(G^2)$.

To reiterate, we have obtained a new balance law,
\begin{gather}
	\J^+ - \J^- = - \Delta_\J,
	\nonumber\\
	\Delta_\J^{\mathstrut} \equiv \Delta_\J^\rad + \Delta_\J^\stat,
\label{eq:balance_mech_J}
\end{gather}
which equates the total loss of mechanical angular momentum to the sum of a radiative term~$\smash{\Delta_\J^\rad}$ and a static term~$\smash{\Delta_\J^\stat}$ [see~Eqs.~\eqref{eq:loss_mech_J_rad_Lorentz} and~\eqref{eq:loss_mech_J_stat_Lorentz} for the Lorentz-tensor versions of these quantities]. The former admits the interpretation of being the total amount of angular momentum carried away from the binary by radiation, whereas the latter accounts for the fact that angular momentum can also be deposited into the static components of the gravitational field. The result in Eq.~\eqref{eq:balance_mech_J} is accurate to all orders in~$G$, but when a post-Minkowskian expansion is performed, one finds that the radiative term always starts at~$O(G^3)$, while the static term always~starts~at~$O(G^2)$.

\subsection{Supertranslation invariance}
\label{sec:mech_supertranslation}

The first of two puzzles discussed in the Introduction raised the question as to whether the $O(G^2)$ part of the Bondi flux~$F_J$ is physical, given that it can be removed by a supertranslation. As we have just shown, $F_J$ does not balance the loss of mechanical angular momentum from the binary. Instead, the relevant quantity is ${\Delta_\J^{\mathstrut} \equiv \Delta_\J^\rad + \Delta_\J^\stat}$. Here we establish that both ${ \Delta_\J^\rad }$ and ${ \Delta_\J^\stat }$ are inherently physical by showing that they are individually invariant under pure~supertranslations.

Consider two Bondi frames $(u,\theta^A)$ and $(u',\theta^A)$ that are related by the pure supertranslation ${u' = u - \alpha_{\ell\geq 2}}$. Under this transformation, the initial and final values of the shear transform~as
\begin{equation}
	C^{\prime\pm}(\theta^A) = C^\pm(\theta^A) + \alpha_{\ell\geq 2}(\theta^A).
\label{eq:transformation_C}
\end{equation}
Because $\smash{\dot M = 0}$ on~$\scri^+_\pm$, we know that $M^-$ and $M^+$ are both invariant under this transformation, as is~$\Delta C$. The shift $\Delta Z$, and consequently the quantity~${\Delta S = \Delta C + \Delta Z}$, are also unaffected. Meanwhile, the dynamical part of the shear tensor and the news tensor transform as~\cite{Javadinezhad:2018urv}
\begin{subequations}
\begin{align}
	\dC_{AB}'(u', \theta^C) &= \dC_{AB}(u' + \alpha_{\ell\geq 2}(\theta^C),\theta^C),
	\\
	N_{AB}'(u', \theta^C) &= N_{AB}(u' + \alpha_{\ell\geq 2}(\theta^C),\theta^C).
\end{align}
\end{subequations}

It now follows from these transformation rules that $\smash{ \Delta_\J^\stat }$ is manifestly invariant under pure supertranslations, and since the Bondi angular momenta~$J^\pm$ are known to transform as ${J^{\pm\prime} = J^\pm + j(M^\pm, \alpha_{\ell\geq 2})}$~\cite{Flanagan:2015pxa}, it follows from Eq.~\eqref{eq:def_mech_J_binary} and the above that the mechanical angular momenta~$\J^\pm$ are also invariant under pure supertranslations. This immediately implies that $\smash{ \Delta_\J^\rad }$ is also invariant as a consequence of Eq.~\eqref{eq:balance_mech_J}, but it is nevertheless instructive to verify this~explicitly.

Our approach is to recognize that this radiative flux can be rewritten in terms of the two functions
\begin{subequations}
\label{eq:def_invariant_tensors}
\begin{align}
	\iC_{AB}(u, \theta^C) &\coloneq \dC_{AB}(u - C^-(\theta^C), \theta^C),
	\\
	\iN_{AB}(u, \theta^C) &\coloneq N_{AB}(u - C^-(\theta^C), \theta^C),
\end{align}
\end{subequations}
which we shall call the ``invariant shear tensor'' and ``invariant news tensor,'' respectively, on account of the fact that they are invariant under pure supertranslations, i.e.,
\begin{subequations}
\label{eq:invariant_tensors_rule}
\begin{align}
	\iC'_{AB}(u,\theta^C) &= \iC_{AB}^{\vphantom{\prime}}(u,\theta^C),
	\\
	\iN'_{AB}(u,\theta^C) &= \iN_{AB}^{\vphantom{\prime}}(u,\theta^C).
\end{align}
\end{subequations}

To rewrite Eq.~\eqref{eq:loss_mech_J_rad} in terms of these objects, we perform a change of integration variable by replacing ${u \mapsto u - C^-}$, under which
\begin{equation}
	\dC_{AB}(u,\theta^C) \mapsto \dC_{AB}(u - C^-,\theta^C)	= \iC_{AB}(u,\theta^C),
\end{equation}
and likewise for $N_{AB}$; the equality follows from the definition in Eq.~\eqref{eq:def_invariant_tensors}. Caution must be exercised when a covariant derivative acts on one of these tensors, however. For a term like $D_A \dC_{BC}(u, \theta^D)$, the derivative acts only on the angular arguments of $\dC_{BC}$ prior to the change of variable.~This behavior must be preserved after the fact;~hence,%
\begin{align}
	D_A \dC_{BC}(u)
	\mapsto &\;
	D_A \dC_{BC}(u - C^-)
	+
	N_{BC}(u - C^-) D_A C^-
	\nonumber\\
	&=
	D_A \iC_{BC}(u) + \iN_{BC}(u) D_A C^-,
\end{align}
where we have suppressed the dependence on~$\smash{\theta^A}$ for readability. In the first line, the first term on the rhs has $D_A$ acting on all three arguments of~$\dC_{BC}$; the effect of $D_A$ acting on the first argument ${u - C^-}$ is cancelled by the second term. The second line then follows~from~Eq.~\eqref{eq:def_invariant_tensors}.

After making this change of variables, we find that the explicit dependence on $C^-$ drops out, and we are left with
\begin{align}
	\Delta_\J^\rad
	=&
	\int\frac{\dx u \dx^2\Omega}{32\pi G} Y^A
	\bigg(
		\iN^{BC} D_A \iC_{BC}
		-
		2D_B(\iN^{BC} \iC_{AC})
		\nonumber\\&
		+
		\frac{1}{2} D_A(\iN^{BC} \iC_{BC})
		-
		\frac{1}{2} u D_A(\iN^{BC} \iN_{BC})
	\bigg).
\label{eq:loss_mech_J_flux_inv}
\end{align}
It is interesting to see that this expression is structurally identical to the Bondi flux~$F_J$ in Eq.~\eqref{eq:flux_J}, except that the invariant tensors $\{ \iC_{AB}, \iN_{AB} \}$ have assumed the role~of $\{ C_{AB}, N_{AB} \}$.~Indeed, specializing to the canonical frame wherein ${C^- = 0}$ would make them equivalent to one another. This is as it should be, since it was already understood in Ref.~\cite{Veneziano:2022zwh} (see also Ref.~\cite{Javadinezhad:2022ldc}) that the Bondi flux~$F_J$ gives precisely the radiated flux when computed in a canonical frame. Our Eq.~\eqref{eq:loss_mech_J_flux_inv} [or~equivalently,~Eq.~\eqref{eq:loss_mech_J_rad}] generalizes the result of Ref.~\cite{Veneziano:2022zwh} by providing an expression for the radiated flux that holds in any Bondi frame. With~that~said, we reiterate that neither $F_{J}$ nor $\smash{\Delta_\J^\rad}$ give the total loss of mechanical angular momentum from the binary, as one must still account for the additional contribution from~$\smash{\Delta_\J^\stat}$.

Returning to the issue of supertranslation invariance, we have thus far shown that Eq.~\eqref{eq:loss_mech_J_flux_inv} is an alternative but equivalent way of writing Eq.~\eqref{eq:loss_mech_J_rad}. To~complete the proof, let $\smash{\Delta_\J^\rad}$ and $\smash{\Delta_\J^{\rad\prime}}$ be the total fluxes across~$\scri^+$ as measured in the two frames $(u,\theta^A)$ and $(u',\theta^A)$. The flux $\smash{\Delta_\J^\rad}$ in the unprimed frame is given by Eq.~\eqref{eq:loss_mech_J_flux_inv}, but we can replace $\smash{\iC_{AB}^{\mathstrut} \mapsto \iC'_{AB}}$ and $\smash{\iN_{AB}^{\mathstrut} \mapsto \iN'_{AB}}$ without issue as a consequence of Eq.~\eqref{eq:invariant_tensors_rule}. Then renaming the integration variable $u$~to~$u'$ shows~that~$\smash{ \Delta_\J^{\rad} = \Delta_\J^{\rad\prime} }$.

It is worth remarking here that other studies have also recently sought to propose definitions of the angular momentum that are invariant under (pure) supertranslations \cite{Compere:2019gft, Chen:2021szm, Chen:2021kug, Mao:2023evc, Javadinezhad:2022hhl, Javadinezhad:2022ldc}.~In fact, what we call the mechanical angular momentum $\J$ is closely related to the definition proposed in Refs.~\cite{Compere:2019gft, Chen:2021szm, Chen:2021kug, Mao:2023evc}, except that the latter does not include the translation~$Z^+$ that corrects for the shift in reference point. The consequence of this, and a separate proposal~from~Ref.~\cite{Javadinezhad:2022ldc}, are discussed in more detail~in~Appendix~\ref{app:comparisons}.

To close this section, we shall briefly consider how $\Delta_\J$ transforms under the remainder of the BMS~group. Under a translation ${u' = u - \alpha_{\ell\leq 1}}$, for instance, we have~that ${C' = C}$, and thus $\smash{ D_A \iC'\!{}_{BC}(u') = D_A \iC_{BC}(u) + \iN_{BC}(u) D_A \alpha_{\ell\leq 1} }$. This can be used to show that
\begin{equation}
	\Delta_\J^{\rad\prime} = \Delta_\J^\rad
	+
	\int \frac{\dx u}{8G} j(\iN^{AB} \iN_{AB}, \alpha_{\ell\leq 1}).
\end{equation}
It then follows from Eq.~\eqref{eq:def_invariant_tensors} and the freedom to change integration variables that we can substitute $\iN^{AB} \iN_{AB}$ for $N^{AB} N_{AB}$ in the above without issue.~Now parametrizing ${\alpha_{\ell\leq 1} = (n \cdot a)}$ by the constant vector~$a^\mu$ and then switching to the Lorentz-tensor representation (the steps are almost identical to those in Appendix~\ref{app:Lorentz_mech_J}), we~get
\begin{equation}
	\Delta_{\J\rad}^{\prime\mu\nu} = \Delta_{\J\rad}^{\mu\nu} + 2 a^{[\mu}_{\mathstrut} F_P^{\nu]}.
	\label{eq:calFJ_translations}
\end{equation}
The static term $\smash{\Delta_\J^\stat}$ is also invariant under translations because the shift $\Delta Z$ is unchanged when $\J^+$ and $\J^-$ are transformed by the same amount~$a^\mu$; hence,
\begin{align}
	\Delta_\J^{\prime\mu\nu} &= \Delta_\J^{\mu\nu} + 2 a^{[\mu}_{\mathstrut} F_P^{\nu]},
	\\
	\J_\pm^{\prime\mu\nu} &= \J_\pm^{\mu\nu} + 2 a^{[\mu}_{\mathstrut} P_\pm^{\nu]},
\end{align}
as one should expect from the balance laws in Eqs.~\eqref{eq:balance_P_Lorentz} and \eqref{eq:balance_mech_J}; note that the four-momenta $P_\pm^\mu$ and the corresponding flux $F_P^\mu$ are invariant under both translations and pure supertranslations~\cite{Flanagan:2015pxa}. Deriving the remaining transformation rules under the Lorentz group is considerably more involved, but the results in, e.g., Refs.~\cite{Flanagan:2015pxa, Javadinezhad:2022hhl} can be used to show that these objects are all indeed covariant under Lorentz transformations.

% ---------------------------------------------------------- %
\section{Post-Minkowskian results}
\label{sec:pm}

This section computes the total loss of mechanical angular momentum during a two-body scattering encounter at leading order in the post-Minkowskian expansion, i.e., at~$O(G^2)$. Because the radiative flux~$\smash{\Delta_\J^\rad}$ begins only at~$O(G^3)$, the total loss~$\Delta_\J$ is determined solely by the static term~$\smash{\Delta_\J^\stat}$ at leading order. We compute~$\smash{\Delta_\J^\stat}$ explicitly at~$O(G^2)$ in Sec.~\ref{sec:pm_stat} and find that it agrees with the result obtained from quantum field theory~\cite{Manohar:2022dea, DiVecchia:2022owy, DiVecchia:2022piu} in all Bondi frames. In fact, we find that our expression for $\smash{\Delta_\J^\stat}$ agrees with the static part of the result in Refs.~\cite{DiVecchia:2022owy, DiVecchia:2022piu} also at~$O(G^3)$, although we omit the lengthy details in this case. In~Sec.~\ref{sec:pm_cm}, we explain the connection between $\smash{\Delta_\J^\stat}$ and the Bondi flux~$F_J$, and thus why their space-space components just so happen to agree at~$O(G^2)$ in the binary's \cm~frame.

\subsection{Static contribution}
\label{sec:pm_stat}

Three quantities are needed to determine $\smash{\Delta_\J^\stat}$ in Eq.~\eqref{eq:loss_mech_J_stat}: the final value of the mass aspect~$M^+$, the gravitational memory~$\Delta C$, and the shift~$\Delta Z$. In principle, $M^+$ can be obtained purely from the information on~$\scri^+$ by solving the Einstein equation in Eq.~\eqref{eq:Einstein_M} once we are given the initial condition~$M^-$ and the news tensor~$N_{AB}$; in practice, however, it is easier to solve the equations of motion for the trajectories of the two bodies directly~\cite{doi:10.1088/0305-4470/13/12/017, Damour:2016gwp, Bern:2019crd, Kalin:2020mvi}. Their final four-momenta~$p_a^+$ can then be plugged into Eq.~\eqref{eq:twobody_M_initial_and_final} to~give~$M^+$. For~${\Delta S \equiv \Delta C + \Delta Z}$, we combine Eqs.~\eqref{eq:twobody_Z_C_initial} and~\eqref{eq:twobody_Z_C_final} to~obtain
\begin{equation}
	\Delta S
	=
	\sum_{a=1}^2
	\bigg[
		2G (n\cdot p_a) \log\left(\frac{-n\cdot p_a}{m_a}\right)
	\bigg]^{+\infty}_{-\infty}
	+
	O(G\Delta\mathcal{E}).
\label{eq:PM_solution_Delta_S}
\end{equation}

It is possible to check that our result for ${\Delta C \equiv \mathbb{P}_{\ell\geq 2}\Delta S}$ is consistent with the Einstein equations. First differentiate Eq.~\eqref{eq:twobody_CAB} twice and then use the identity in Eq.~\eqref{eq:commutator}~to~show~that
\begin{equation}
	-D^2 (D^2+2) \Delta C
	=
	D^A D^B \Delta C_{AB}
	=
	4G(\Delta M + \Delta\mathcal{E}),
\label{eq:Einstein_Delta_C}
\end{equation}
where the second equality follows from using Eq.~\eqref{eq:DeltaM}, and recall that $\Delta\mathcal{E}$ is the total energy radiated per unit solid angle across~$\scri^+$.~The differential operator ${D^2(D^2 + 2)}$ is invertible via the method of Green functions~\cite{Bieri:2013ada, Pasterski:2015tva}, and so formally the solution~is
\begin{equation}
	\Delta C
	=
	-4G [D^2 (D^2+2)]^{-1}
	(\Delta M + \Delta\mathcal{E}) .
\label{eq:formal_solution_Delta_C}
\end{equation} 
The first term involving~$\Delta M$ is known as the ``linear memory'' and depends only on the initial and final momenta of the two bodies~\cite{Zeldovich:1974gvh, doi:10.1038/327123a0}; the second term, which is known as the ``nonlinear memory,'' accounts for the additional contribution from gravitational radiation~\cite{Christodoulou:1991cr, Wiseman:1991ss, Blanchet:1992br, Thorne:1992sdb}.%
\footnote{These two terms are also often called the ``ordinary memory'' and ``null memory,'' respectively \cite{Bieri:2013ada, Tolish:2014oda}.}

The power counting arguments we made in the previous section tell us that $\Delta M$ starts at $O(G)$ while $\Delta\mathcal{E}$ starts at~$O(G^3)$, and so from Eq.~\eqref{eq:formal_solution_Delta_C} we see that the linear and nonlinear parts of the memory start at~$O(G^2)$ and~$O(G^4)$, respectively. Only the former is needed at the order to which we are working, and one can verify by direct substitution that the ${\ell\geq 2}$ harmonics of Eq.~\eqref{eq:PM_solution_Delta_S} are indeed a valid solution to Eq.~\eqref{eq:Einstein_Delta_C}, up to terms associated with the nonlinear memory.

Equation~\eqref{eq:Einstein_Delta_C} does not fix the remaining ${\ell\leq 1}$ harmonics of~$\Delta S$, however, because these modes live in the kernel of the differential operator $D^2(D^2+2)$. Instead, we determined ${\Delta Z \equiv \mathbb{P}_{\ell\leq 1}\Delta S}$ in Sec.~\ref{sec:mech_twobody} by assuming that the leading $1/r$ behavior of the binary spacetime at~$\scri^+_\pm$ is well approximated by the superposition of two boosted Schwarzschild metrics. Tracking how the transformation from harmonic to Bondi coordinates inadvertently shifts the reference point with respect to which the angular momentum is defined then allows us to fix~$\Delta Z$ uniquely.~A~\emph{post hoc} justification for this approach is that it leads to a result for $\smash{ \Delta_\J^\stat }$ that is rightly Lorentz covariant. To~elaborate, we note that because the projection operators $\mathbb{P}_{\ell\leq 1}$ and $\mathbb{P}_{\ell\geq 2}$ do not commute with Lorentz boosts, $j(M^+, \Delta C)$ and $j(M^+, \Delta Z)$ are not individually Lorentz covariant. However, as the quantity ${\Delta S = \Delta C + \Delta Z}$ in Eq.~\eqref{eq:PM_solution_Delta_S} can be written down without the need for these projectors, the sum $\smash{ \Delta_\J^\stat = j(M^+, \Delta C) + j(M^+, \Delta Z)}$ is well behaved under all Lorentz transformations. An~important open question is whether a more systematic procedure exists for determining~$\Delta Z$ to all orders in~$G$ for generic spacetimes, but this is a problem that we shall leave to~the~future.

For now, Eq.~\eqref{eq:PM_solution_Delta_S} will suffice to determine~$\smash{\Delta_\J^\stat}$ up to $O(G\Delta\mathcal{E})$~corrections. To make contact with the results in the post-Minkowskian literature, it is useful here to switch to the Lorentz-tensor representation, and so we shall substitute our expressions for $M^+$ and $\Delta S$ into the formula for ${ \Delta_{\J\stat}^{\mu\nu} }$ as given in Eq.~\eqref{eq:loss_mech_J_stat_Lorentz}. After also using Eq.~\eqref{eq:AD_identities} to evaluate the derivative that acts on~$\Delta S$, we~find~that
\begin{align}
	\Delta_{\J\stat}^{\mu\nu}
	=
	\int & \frac{\dx^2\Omega}{2\pi} \sum_{a=1}^2
	G M^+
	\bigg\{
		2 p_a^{[\mu} n^{\nu]}_{\mathstrut}	
		\bigg[
			1 + \log\bigg(\frac{ -n\cdot p_a}{m_a}\bigg)
		\bigg]
		\nonumber\\&
		+
		(-n\cdot p_a) n^{[\mu}\bar{n}^{\nu]}
	\bigg\}^{+\infty}_{-\infty}
	+ O(G\Delta\mathcal{E}).
\label{eq:memory_2PM}
\end{align}
To complete this calculation, we use the known result for the final four-momenta of the two bodies~\cite{doi:10.1088/0305-4470/13/12/017, Damour:2016gwp, Bern:2019crd, Kalin:2020mvi},
\begin{equation}
	p_{a+}^\mu 
	=
	p_{a-}^\mu 
	+
	(-1)^a
	\frac{2G m_1 m_2 }{|b|^2}
	\frac{2\gamma^2 - 1}{\sqrt{ \gamma^2-1 }}
	b^\mu
	+
	O(G^2),
	\label{eq:impulse_1PM}
\end{equation}
where $\smash{ \gamma \equiv (-p_1^-\cdot p_2^-)/ m_1 m_2 }$ is the Lorentz factor for their initial relative velocity~$v$, the constant vector $b^\mu$ here denotes their impact parameter, and $\smash{ |b| \equiv \sqrt{ b \cdot b} }$.

The integral in Eq.~\eqref{eq:memory_2PM} is challenging to evaluate as is due to its tensor-valued nature, but we can proceed by projecting it along the six independent basis tensors formed by the antisymmetrized outer products of the four basis vectors $\smash{ \{ p_{1-}^\mu, p_{2-}^\mu, b^\mu_{\mathstrut}, \hat{l}^\mu_{\mathstrut} \} }$, where $\hat{l}^\mu$ is the unit spacelike vector orthogonal to all of the other basis vectors. To give an example, the component of ${ \Delta_{\J\stat}^{\mu\nu} }$  along the direction $2 b^{[\mu} \hat{l}^{\nu]}$ is given by $\smash{ (\Delta_{\J\stat}^{\mu\nu} b_\mu \hat{l}_\nu)/|b|^2 }$, which is a Lorentz scalar that we can evaluate in any frame. For convenience, we have chosen the frame in which the second body is initially at rest, i.e., $\smash{ p_{2-}^\mu = (m_2,\mathbf{0})}$, and we have further oriented our spatial axes such that $\smash{ p_{1-}^\mu = \gamma m_1 (1,0,0,v)}$, while $\hat{l}^\mu$ and $b^\mu$ are aligned along the positive $x$ and $y$~directions, respectively.

The end result of this calculation is
\begin{align}
	\Delta_{\J\stat}^{\mu\nu}
	&=
	\frac{2G^2 m_1 m_2}{ |b|^2 }
	\frac{2\gamma^2 - 1}{ \sqrt{\gamma^2-1} }
	\mathcal{I}(\gamma)
	b^{[\mu}_{\mathstrut} (p_{1-}^{\nu]}-p_{2-}^{\nu]})
	+
	O(G^3),
	\nonumber\\
	\mathcal{I}(\gamma)
	&=
	\frac{2(8 - 5\gamma^2)}{3(\gamma^2-1)} + \frac{2\gamma (2\gamma^2 -3)}{(\gamma^2 - 1)^{3/2}} \mathrm{arccosh}(\gamma),
\label{eq:memory_G2}
\end{align}
which agrees with the result obtained from quantum field theory \cite{Manohar:2022dea, DiVecchia:2022owy, DiVecchia:2022piu} in all Bondi frames. In fact, we have used the same steps to evaluate $\smash{\Delta_\J^\stat}$ up to $O(G^3)$, and we again find agreement with the existing literature. This comparison is possible because Refs.~\cite{DiVecchia:2022owy, DiVecchia:2022piu} similarly decompose their formula for the total angular momentum loss into a radiative part and a static part; the latter is what agrees with our~$\smash{\Delta_\J^\stat}$. It will be interesting in the future to verify if our $\smash{\Delta_\J^\rad}$ matches their radiative~part.

\subsection{Center-of-mass frame}
\label{sec:pm_cm}

The second of the two puzzles discussed in the Introduction raised the question as to why there is generally a discrepancy between Refs.~\cite{Manohar:2022dea, DiVecchia:2022owy, DiVecchia:2022piu} (see also Ref.~\cite{Bini:2022wrq}) and Refs.~\cite{Jakobsen:2021smu, Mougiakakos:2021ckm} on the space-space components of the angular momentum loss at~$O(G^2)$, except in the binary's \cm~frame. The previous subsection establishes, at least up to~$O(G^2)$, that the quantity being computed in Refs.~\cite{Manohar:2022dea, DiVecchia:2022owy, DiVecchia:2022piu} is indeed the total loss of mechanical angular momentum~$\Delta_\J$. This suffices to explain why there is \emph{usually} a discrepancy, since what is computed in Refs.~\cite{Jakobsen:2021smu, Mougiakakos:2021ckm} is the Bondi flux~$F_J$, which---as we now know---is not the same as~$\Delta_\J$. It still remains to explain why the space-space components of $F_J$ and $\Delta_\J$ just so happen to agree at $O(G^2)$ in~the~\cm~frame.

Our approach will be to rewrite the static~term~$\smash{\Delta_\J^\stat}$, which recall is the only contribution to $\Delta_\J$ at this order, as the sum of a part that strongly resembles~$F_J$ and another part whose space-space components can be seen to vanish in the \cm~frame. To start with, we introduce ${\hat t^\mu = (1,\mathbf{0})}$ and ${\rv^\mu = n^\mu - \hat t^\mu}$ as the unit vectors in the future-pointing timelike direction and outward-pointing radial direction, respectively. It is then possible to rewrite Eq.~\eqref{eq:twobody_M_initial_and_final}~as
\begin{equation}
	M^\pm = (3 \rv - \hat t) \cdot P^\pm + \frac{1}{4G} D^A D^B f^\pm_{AB}.
\label{eq:def_M_via_f}
\end{equation}
The first term, which depends on the four-momentum of the binary~$P^\pm$, makes up the ${\ell=0}$ and ${\ell=1}$ harmonics of~$M^\pm$, while the remaining harmonics with ${\ell\geq 2}$ are encoded in the symmetric and traceless tensor
\begin{equation}
	f_{AB}^\pm
	=
	4 G
	\bigg(
		\Omega_{AC} \Omega_{BD} - \frac{1}{2}\Omega_{AB} \Omega_{CD}
	\bigg)
	\sum_{a=1}^2
	\frac{ p_{a\pm}^{C} p_{a\pm}^{D}}{(- n\cdot p^\pm_a)}.
\label{eq:def_fAB}
\end{equation}
We define $\smash{ p_{a\pm}^A \equiv e^A_\mu p_{a\pm}^\mu }$ as the projection of $p_{a\pm}^\mu$ along the direction of increasing~$\theta^A$; the properties of the projector~$e^A_\mu$ are described in more detail in Appendix~\ref{app:Lorentz}. For later comparison, we note that an equivalent way of writing Eq.~\eqref{eq:def_fAB}~is
\begin{align}
	f_{AB}^\pm &= -(2 D_A D_B - \Omega_{AB} D^2) f^\pm,
	\nonumber\\
	f^\pm
	&=
	\sum_{a=1}^2 \mathbb{P}_{\ell\geq 2}\!
	\left[
		2G (n\cdot p_a^\pm) \log\left(\frac{-n\cdot p_a^\pm}{m_a}\right)
	\right].
\label{eq:def_fAB_potential}
\end{align}

Now substituting the expression for~$M^+$ in Eq.~\eqref{eq:def_M_via_f} into Eq.~\eqref{eq:loss_mech_J_stat}, using ${\Delta C_{AB} \equiv \int_{-\infty}^{+\infty} N_{AB}\,\dx u}$, and then performing several integrations by parts as described in Appendix~\ref{app:proofs}, we~find~that
\begin{align}
	\Delta_\J^\stat
	&=
	\int  \frac{\dx u \dx^2\Omega}{32\pi G}
	Y^A
	\bigg(
		N^{BC} D_A^{\mathstrut} f^+_{BC}
		-
		2 D_B^{\mathstrut}(N^{BC} f^+_{AC})
		\nonumber\\& \qquad\qquad\;\;
		+
		\frac{1}{2} D_A^{\mathstrut} (N^{BC} f^+_{BC})
	\bigg)
	\nonumber\\
	& +
	\! \int \! \frac{\dx^2\Omega}{8\pi}
	\bigg(
		3 (\rv \cdot P^+)
		(2 Y^C D_C \Delta S - \Delta S D_C Y^C)
		\nonumber\\&\qquad\qquad
		+
		3 (\hat t\cdot P^+) (D_A Y^A) \Delta S
	\bigg).
\label{eq:loss_mech_J_mem_rewrite}
\end{align}

This result is an exact rewriting of Eq.~\eqref{eq:loss_mech_J_stat}, but for our purposes at present, it is safe to neglect terms of order $G^3$ and higher. We can then replace ${ P^+ \mapsto P^- }$ and ${ f^+_{AB} \mapsto f^-_{AB} }$ in Eq.~\eqref{eq:loss_mech_J_mem_rewrite}. Having done so, we see that the terms involving $\smash{ f^-_{AB} }$ are exactly what one would get from computing the Bondi flux~$F_J$ at $O(G^2)$ [cf.~Eq.~\eqref{eq:flux_J}] in the class of intrinsic frames wherein the initial value of the shear tensor ${C^-_{AB} = f^-_{AB}}$ [compare Eq.~\eqref{eq:def_fAB_potential} with Eqs.~\eqref{eq:twobody_C_initial} and~\eqref{eq:twobody_CAB} when the intrinsic gauge ${\beta_{\ell\geq 2}= 0}$ is imposed]; this is exactly what is computed in Refs.~\cite{Jakobsen:2021smu, Mougiakakos:2021ckm}, and~also~Ref.~\cite{Damour:2020tta}.

We select a particular member from this class of intrinsic frames by specifying the ingoing four-momenta~$p_{a-}^\mu$ of the two bodies. Choosing these momenta such that the binary's center of mass is initially at rest sets ${\rv \cdot P^- = 0}$ by definition; hence, in the (intrinsic) \cm~frame, the only difference between $F_J$ and $\smash{\Delta_\J^\stat}$ at $O(G^2)$ is the term involving the initial energy of the binary~${(\hat t \cdot P^-)}$. This term does not contribute to the space-space components of~$\smash{\Delta_\J^\stat}$,~however.

To see this, first note from the last line of Eq.~\eqref{eq:loss_mech_J_mem_rewrite} that this term is proportional to the quantity~$D_A Y^A$, which---as we show below Eq.~\eqref{eq:DBYA}---is equal to ${- \omega_{\mu\nu} n^\mu \bar n^\nu}$ when using the parametrization in Eq.~\eqref{eq:Y_Lorentz}. Since ${ n^\mu \bar n^\nu \equiv 2\rv^{[\mu}\hat{t}^{\nu]} }$, it~follows after differentiating $\smash{\Delta_\J^\stat}$ with respect to $\omega_{\mu\nu}$ that the ${(\hat t \cdot P^-)}$ term contributes only to the time-space components~$\smash{\Delta_{\J\stat}^{0i}}$. This explains why the space-space components of $F_J$ and $\smash{\Delta_\J^\stat}$~fortuitously agree in the (intrinsic) \cm~frame~at~$O(G^2)$.

% ---------------------------------------------------------- %
\section{Conclusion}
\label{sec:conclusion}

We have introduced a new notion of angular momentum for asymptotically flat spacetimes that we call the mechanical angular momentum~$\J$.~This quantity satisfies two key~properties.~First, we showed by considering the example of a boosted Schwarzschild spacetime that it is the mechanical angular momentum~$\J$ that depends only on the trajectory and four-momentum of the black hole.~This is in contrast to the standard Bondi angular momentum~$J$, which is equal to the sum of $\J$ and an extra piece involving the shear of the gravitational field~$C$. Second, we showed that---also unlike~$J$---$\J$~is invariant under pure supertranslations.

We then derived a new balance law that is written explicitly in terms of~$\J$. In~doing~so, we found that the total loss of mechanical angular momentum $\smash{ \Delta_\J }$ naturally splits into the sum of two terms: a radiative term~$\smash{ \Delta_\J^\rad }$ [Eq.~\eqref{eq:loss_mech_J_rad}], which describes the transfer of angular momentum into radiation, and a static term~$\smash{ \Delta_\J^\stat }$ [Eq.~\eqref{eq:loss_mech_J_stat}], which accounts for the fact that angular momentum can also be deposited into the static components of the gravitational field. Both terms are inherently physical, as we showed that they are individually invariant under pure~supertranslations.

Interestingly, our definition for~$\J$ bears a strong resemblance to other recent proposals for a supertranslation-invariant version of the angular momentum~\cite{Compere:2019gft, Chen:2021szm, Chen:2021kug, Mao:2023evc, Javadinezhad:2022hhl}, and in fact, all of these definitions coincide for the initial state of the binary at~$\scri^+_-$. The key novelty in our definition is the addition of a translation~$\Delta Z$ in the final state at~$\scri^+_+$. Per the discussion below Eq.~\eqref{eq:twobody_Z_final}, we interpreted this term as correcting for an inadvertent shift in the reference point about which the angular momentum is defined. It remains an open question as to how $\Delta Z$ should be determined for generic spacetimes, but for the case of two-body scattering that is the main focus of this work, reasonable assumptions about the system in the asymptotic past and future were sufficient to fix $\Delta Z$ uniquely---at least, at the first few orders in the post-Minkowskian expansion. As a kind of \emph{post hoc} justification, we also found that the inclusion of $\Delta Z$ is essential if $\smash{ \Delta_\J^\stat }$ is to be Lorentz~covariant.

Our formula for the mechanical angular momentum loss~$\smash{ \Delta_\J^{\mathstrut} \equiv \Delta_\J^\rad + \Delta_\J^\stat }$ is accurate to all orders in~$G$, but when a post-Minkowskian expansion is performed, one finds that $\smash{ \Delta_\J^\stat }$ always starts at~$O(G^2)$, while $\smash{ \Delta_\J^\rad }$ starts only later at~$O(G^3)$. At~$O(G^2)$, we were able to explain why the space-space components of the Bondi angular momentum flux $F_J$ just so happen to give the same result as that of $\Delta_\J$ in the binary's (instrinsic) \cm~frame, and thus why previous calculations utilizing the former obtained the correct result. We also showed how to compute $\Delta_\J$ explicitly at~$O(G^2)$, which is in agreement with the results in Refs.~\cite{Damour:2020tta, Manohar:2022dea, DiVecchia:2022owy, DiVecchia:2022piu}. Moreover, we have verified that $\smash{ \Delta_\J^\stat }$ matches the corresponding static part of the result in Refs.~\cite{DiVecchia:2022owy, DiVecchia:2022piu} also at~$O(G^3)$. That these results are all in agreement establishes a clearer link between the notions of angular momentum used in these quantum field theoretic approaches to the two-body problem, on the one hand, and that of classical general relativity, on~the~other.

In the future, it will be interesting to refine this connection by verifying that our total loss $\Delta_\J$ matches the results of Refs.~\cite{Manohar:2022dea, DiVecchia:2022owy, DiVecchia:2022piu} at~$O(G^3)$. Computing $\smash{ \Delta_\J^\rad }$ would require knowledge of the waveform ${\dC_{AB} \equiv \dC_{AB}(u,\theta^C)}$ and its first derivative~${ N_{AB} \equiv \partial_u \dC_{AB} }$ up to~$O(G^2)$, but in practice, we know that a direct evaluation of the position-space integral in Eq.~\eqref{eq:loss_mech_J_rad} is too challenging for existing methods. Even the simpler integral for the four-momentum flux is prohibitively difficult in position space \cite{Jakobsen:2021smu, Mougiakakos:2021ckm}, which is why exact results have mostly been obtained via a momentum-space integral involving the square of the amplitude for on-shell graviton emission~\cite{Herrmann:2021lqe, Herrmann:2021tct, DiVecchia:2021bdo, Bjerrum-Bohr:2021din, Riva:2021vnj, Mougiakakos:2022sic, Jakobsen:2022psy, Heissenberg:2022tsn, Riva:2022fru, Jakobsen:2022zsx}. The key question, then, is how to relate the Fourier transforms of $\dC_{AB}$ and~$N_{AB}$ to the aforementioned amplitude; the former are objects defined in Bondi gauge, whereas the latter is usually computed in de~Donder gauge, and the transformation between the two is not trivial~\cite{Isaacson:1968zz, Blanchet:1986dk, Blanchet:2020ngx}. 

The computation of~$\Delta_\J$ at $O(G^4)$ will also be particularly interesting, because it is at this order that the nonlinear part of the gravitational memory first contributes, as can be seen from Eqs.~\eqref{eq:loss_mech_J_stat} and~\eqref{eq:formal_solution_Delta_C}. Presently, no result for the angular momentum loss at this order has been obtained via any approach, and so it will be interesting to see if ours continues to make the same predictions as those based on quantum field theory. On a more fundamental level, it will also be interesting to gain a deeper understanding of how this work sits in relation to the wider web of connections that have been drawn between asymptotic symmetries, soft theorems, and memory effects~\cite{Strominger:2013lka, He:2014laa, Strominger:2014pwa, Campiglia:2015kxa, Pasterski:2015tva, Strominger:2017zoo, Cristofoli:2022phh}.
\\

\paragraph{Note added}
The main change in this v2, compared to the original analysis of v1, is the inclusion of the translation $\Delta Z$ at $\scri^+_+$, which is crucial for ensuring the Lorentz invariance of $\smash{\Delta_\J^\stat}$. We obtained $\Delta Z$ by considering the coordinate transformation from harmonic to Bondi coordinates, as detailed in Appendix~\ref{app:schwz} (see also Ref.~\cite{Veneziano:2022zwh}). To sharpen the focus of this paper, we have also removed a discussion on the precise form of the mechanical angular momentum~$\J$ at $\scri^+_\pm$, since only the relation between $\J^\pm$ and $J^\pm$ [Eq.~\eqref{eq:def_mech_J_binary}] is needed to determine the total angular momentum loss~$\Delta_\J$. The contents of Appendix~B of~v1, which entered into these details, have therefore been removed.

\acknowledgments
It is a pleasure to thank Reza Javadinezhad, Massimo Porrati,  Rodolfo Russo, and Gabriele Veneziano for insightful discussions, and Thibault Damour and Eanna Flanagan  for helpful comments on a previous version of this manuscript.
M.M.R. is funded by the Deutsche Forschungsgemeinschaft (DFG, German Research Foundation) under Germany's Excellence Strategy -- EXC 2121 ``Quantum Universe'' -- 390833306.
M.M.R. would also like to acknowledge the kind hospitality of New York University and Yale University, where some of this work was completed.
This work was partially supported by the Centre National d'\'{E}tudes Spatiales~(CNES).
The xAct package~\cite{xAct} for Mathematica was used to aid some of our calculations.

% ---------------------------------------------------------- %
\appendix
\section{Lorentz tensors}
\label{app:Lorentz}

This appendix provides a dictionary for converting between the scalar-valued integrals $\{ P(\sigma), J(\sigma), \dots \}$ of the Bondi-Sachs formalism and the Lorentz-tensor representation $\smash{ \{ P^\mu_{\mathstrut}, J^{\mu\nu}_{\mathstrut}, \dots \} }$ of the charges and their corresponding fluxes. We begin by developing the required mathematical machinery in Sec.~\ref{app:Lorentz_basis}. To illustrate the general principles of its use, we prove the result of Eq.~\eqref{eq:def_mech_J_schwz_Lorentz} for the mechanical angular momentum $\smash{\J^{\mu\nu}}$ of a boosted Schwarzschild spacetime in Sec.~\ref{app:Lorentz_mech_J}. Explicit expressions for the fluxes, valid for any spacetime, are then presented in Sec.~\ref{app:Lorentz_fluxes}.

\subsection{Basis vectors}
\label{app:Lorentz_basis}

In Sec.~\ref{sec:bondi_BMS} of the main text, we introduced the Lorentzian coordinates ${x^\mu \equiv (t,x,y,z)}$ and two null vectors, $n^\mu$~and~$\bar{n}^\mu$. Two more spacelike vectors are needed to form a basis that spans Minkowski space. We~introduce
\begin{equation}
	e_A^\mu = \frac{\partial n^\mu}{\partial\theta^A}
	\quad
	(\theta^A \in \{\theta,\phi\})
\label{eq:e_A}
\end{equation}
as the two basis vectors tangent to the unit 2-sphere. Together, our four basis vectors satisfy
\begin{align}
	&
	n_\mu \bar{n}^\mu = -2,
	\quad
	\eta_{\mu\nu} e^\mu_A e^{\nu \vphantom{\mu}}_B = \Omega_{AB},
	\nonumber\\
	&
	n_\mu n^\mu
	=
	\bar{n}_\mu \bar{n}^\mu
	=
	n_\mu e^\mu_A
	=
	\bar{n}_\mu e^\mu_A
	= 0,
\label{eq:minkowski_basis_constraints}
\end{align}
where indices are always raised and lowered with the two metrics $\eta_{\mu\nu}$~and~$\Omega_{AB}$.

The vectors $e^\mu_A$ and their duals also allow us to map a tensor defined on the 2-sphere onto Minkowski space; for instance, the Lorentzian version of the shear tensor~$C_{AB}$ is ${ C_{\mu\nu} \equiv e_\mu^A e_{\nu \vphantom{\mu}}^B C_{AB} }$. Another important tensor that we will need is the induced~metric
\begin{equation}
	\Omega_{\mu\nu}
	=
	e_\mu^A e_{\nu \vphantom{\mu}}^B \Omega_{AB}
	=
	\eta_{\mu\nu} + n_{(\mu} \bar{n}_{\nu)},
\label{eq:def_induced_metric}
\end{equation}
where the last equality follows from Eq.~\eqref{eq:minkowski_basis_constraints}. This map is, of course, not always invertible, because a general tensor $X^{\mu_1 \cdots \mu_n}$ can also have components that are orthogonal to~$\smash{ e^\mu_A }$. For these objects, it is useful to introduce the transverse projection
\begin{equation}
	[X^{\mu_1 \cdots \mu_n}]^\text{T}
	\coloneq
	\Omega^{\mu_1}{}_{\nu_1} \cdots \Omega^{\mu_n}{}_{\nu_n} X^{\nu_1 \cdots \nu_n}, 
\end{equation}
which can then be readily pulled back onto the 2-sphere. As~a piece of terminology, we shall say that a tensor $A^{\mu_1 \cdots \mu_n}$ is transverse if~$\smash{ A^{\mu_1 \cdots \mu_n} = [A^{\mu_1 \cdots \mu_n}]^\text{T} }$.

With these definitions in hand, we are now in a position to map the covariant derivative $D_A$ onto Minkowski space. Its counterpart is the angular partial derivative operator,
\begin{equation}
	\eth_\mu 
	\coloneq
	e_\mu^A \frac{\partial}{\partial\theta^A}
	\equiv
	r \Omega_\mu{}^\nu \frac{\partial}{\partial x^\nu}.
\label{eq:def_AD}
\end{equation}
To see how the two derivative operators $D_A$ and $\eth_\mu$ are connected, consider the action of $\eth_\nu$ on the transverse vector~$A^\mu$. We find that
\begin{align}
	\eth_\nu A^\mu
	&= 	
	e_{\nu\mathstrut}^B \partial_B^{\mathstrut} (e^\mu_A A^A)
	\nonumber\\
	&=
	e_{\nu\mathstrut}^B e^\mu_A (\partial_B^{\mathstrut} A^A)
	+
	e_{\nu\mathstrut}^B A^A (\partial_B^{\mathstrut} e^\mu_A).
\label{eq:AD_expand}
\end{align}
Direct evaluation reveals that
$\smash{
	\partial_B^{\mathstrut} e^\mu_A
	=
	e^\mu_C \Gamma^C{}^{\mathstrut}_{\! AB}
	-
	\rv^\mu_{\mathstrut} \Omega^{\mathstrut}_{AB}
}$,
where $\Gamma^C{}_{AB}$ is the Levi--Civita connection on $\Omega_{AB}$, and recall that ${\rv^\mu = (n^\mu - \bar{n}^\mu)/2}$ is the unit spacelike vector pointing in the outward radial direction. Substituting this back into Eq.~\eqref{eq:AD_expand} and using
$\smash{ D_B A^A = \partial_B A^A  + \Gamma^A{}_{BC} A^C }$, we~then~obtain
\begin{equation}
	\eth_\nu A^\mu 
	=
	e^B_{\nu\mathstrut} e^{\mathstrut\mu}_A (D_B A^A) - \rv^\mu A_\nu.
\end{equation}
This result tells us that even if $A^\mu$ is transverse, $\eth_\nu A^\mu$ can have a component that is not tangent to the 2-sphere. It is nevertheless straightforward to project this unwanted component away; we have that $\smash{ [\eth_\nu A^\mu]^\text{T} = e^B_{\nu\mathstrut} e^{\mathstrut\mu}_A (D_B A^A) }$, and more generally
\begin{equation}
	[\partial_\alpha A^{\mu \cdots}{}_{\nu \cdots}]^\text{T}
	=
	(e_\alpha^C e^{\mu}_{A} e^B_{\nu\mathstrut} \cdots) (D_C A^{A \cdots}{}_{B \cdots }). 
\end{equation}

Two useful identities involving~$\eth_\mu$ are
\begin{subequations}
\label{eq:AD_identities}
\begin{gather}
	\eth_\mu n_\nu = -\eth_\mu \bar{n}_\nu = \Omega_{\mu\nu},
	\label{eq:AD_identities_n}
	\\
	\eth_\alpha \Omega_{\mu\nu} = -2 \rv_{(\mu} \Omega_{\nu)\alpha}.
	\label{eq:AD_identities_metric}
\end{gather}
\end{subequations}
The first line follows directly from Eqs.~\eqref{eq:e_A} and~\eqref{eq:minkowski_basis_constraints}, whereas the second follows from Eqs.~\eqref{eq:def_induced_metric} and~\eqref{eq:AD_identities_n}.

\subsection{Mechanical angular momentum}
\label{app:Lorentz_mech_J}

To illustrate how this map is used, here we prove the result in Eq.~\eqref{eq:def_mech_J_schwz_Lorentz} for the mechanical angular momentum ${\J \equiv j(M,B)}$ of a boosted Schwarzschild black hole. Our starting point is the definition for $j(M,B)$ in Eq.~\eqref{eq:def_little_j}, which becomes
\begin{equation}
	j(M,B)
	=
	\int\frac{\dx^2\Omega}{8\pi}
	M (2 Y^A D_A B - B D_A Y^A)
\label{eq:Lorentz_mech_J_raw}
\end{equation}
after an integration by parts. Next, we convert the directional derivative $Y^A D_A$ and the scalar quantity~$D_A Y^A$ into their Lorentz-tensor counterparts. By combining the parametrization of~$Y^A$ in Eq.~\eqref{eq:Y_Lorentz} with the definition provided in~Eq.~\eqref{eq:e_A}, we see that
\begin{equation}
	Y_A = -\omega_{\mu\nu} n^\mu e_A^\nu
\label{eq:Y_Lorentz_eA}.
\end{equation}
Contracting this with $\Omega^{AB}\partial_B$ and using Eq.~\eqref{eq:def_AD} then yields
\begin{equation}
	Y^A \partial_A = - \omega_{\mu\nu} n^\mu \eth^\nu.
\end{equation}
This will suffice for our purposes, as the derivative operator $Y^A D_A$ in Eq.~\eqref{eq:Lorentz_mech_J_raw} acts only on the scalar function~$B$. To get an expression for~$D_A Y^A$, we instead contract Eq.~\eqref{eq:Y_Lorentz_eA} with $e_\alpha^A$ to obtain ${Y_\alpha = -\omega_{\mu\nu} n^\mu \Omega^{\nu}{}_{\alpha} }$. We can use this to show~that
\begin{align}
	D_B Y_A
	&=
	e_B^\rho e_A^\alpha (\eth_\rho Y_\alpha)
	\nonumber\\
	&=
	-\omega_{\mu\nu} e_B^\rho e^\alpha_A
	\bigg(
		\Omega^\mu{}_\rho \Omega^\nu{}_\alpha
		+
		\frac{1}{2} n^\mu \bar{n}^\nu \Omega_{\rho\alpha}
	\bigg),
\label{eq:DBYA}
\end{align}
where the second line follows from Eqs.~\eqref{eq:minkowski_basis_constraints} and~\eqref{eq:AD_identities}. Contracting with $\Omega^{AB}$ then yields $\smash{ D_A Y^A = -\omega_{\mu\nu} n^\mu \bar{n}^\nu }$
as a special case. These results allow us to write
\begin{equation}
	j(M,B)
	=
	\int\frac{\dx^2\Omega}{8\pi}
	\omega_{\mu\nu} M
	\big(
		{-}2 n^\mu \eth^\nu B
		+
		B n^\mu \bar{n}^\nu
	\big).
\end{equation}

Now using ${B = (n \cdot b)}$, we see that the integrand
\begin{align}
	&
	\omega_{\mu\nu} M
	\big(
		{-}2 n^\mu \eth^\nu B
		+
		B n^\mu \bar{n}^\nu
	\big)
	\nonumber\\
	&=
	\omega_{\mu\nu} M b_\rho
	\big(
		{-}2 n^\mu \Omega^{\nu\rho}
		+
		n^\rho n^\mu \bar{n}^\nu
	\big)
	\nonumber\\
	&=
	2 \omega_{\mu\nu} M b^\mu n^\nu.
\end{align}
The second line follows from Eq.~\eqref{eq:AD_identities}, while the third follows from Eq.~\eqref{eq:def_induced_metric}. Putting everything together, we obtain
\begin{equation}
	j(M,B)
	=
	\omega_{\mu\nu} b^\mu
	\bigg(
		\int\frac{\dx^2\Omega}{4\pi} M n^\nu
	\bigg).
\end{equation}
The integral in parentheses gives the four-momentum~$p^\nu$ of the black hole. Differentiating with respect to~$\omega_{\mu\nu}$ and using
$\smash{
	\partial \omega_{\rho\sigma}/\partial\omega_{\mu\nu}
	=
	2 \delta^{[\mu}_\rho \delta^{\nu]}_\sigma
}$
then returns the result in Eq.~\eqref{eq:def_mech_J_schwz_Lorentz}.

\subsection{Flux formulas}
\label{app:Lorentz_fluxes}

Here we present explicit expressions for the various fluxes in their Lorentz-tensor form. These results all follow from a straightforward application of the identities derived in earlier parts of this appendix. First, Eqs.~\eqref{eq:flux_P} and~\eqref{eq:Lorentz_fluxes} give us the four-momentum flux%
\footnote{Notice that the term in Eq.~\eqref{eq:flux_P} that is linear in the news tensor does not contribute to $F_P^\mu$ as it vanishes after an integration by parts; it contributes only to the flux of supermomentum.}
\begin{equation}
	F_P^\mu
	=
	\int \frac{\dx u\dx^2\Omega}{32\pi G}
	(N^{\rho\sigma} N_{\rho\sigma}) n^\mu.
\label{eq:flux_P_Lorentz}
\end{equation}
From Eqs.~\eqref{eq:flux_J} and~\eqref{eq:Lorentz_fluxes}, we obtain the Bondi angular momentum flux
\begin{align} 
	F_J^{\mu\nu}
	=
	\int \frac{\dx u\dx^2\Omega}{32\pi G}
	\big[&
		4 C^{\rho[\mu}_{\mathstrut} N^{\nu]}_{\mathstrut}{}_\rho
		-
		2 N^{\rho\sigma} n^{[\mu}\eth^{\nu]} C_{\rho\sigma}
		\nonumber\\[-0.5em]&
		-
		n^{[\mu} \bar{n}^{\nu]}
		N^{\rho\sigma} \partial_u( u C_{\rho\sigma})
	\big].
\label{eq:flux_J_Lorentz}
\end{align}
For the total mechanical angular momentum loss, the radiative term in Eq.~\eqref{eq:loss_mech_J_rad}~becomes
\begin{align}
	\Delta_{\J\rad}^{\mu\nu}
	=
	\int & \frac{\dx u\dx^2\Omega}{32\pi G}
	\big[
		4 \dC^{\rho[\mu}_{\mathstrut} N^{\nu]}_{\mathstrut}{}_\rho
		-
		2 N^{\rho\sigma} n^{[\mu} \eth^{\nu]} \dC_{\rho\sigma}
		\nonumber\\&
		-
		n^{[\mu} \bar{n}^{\nu]}
		N^{\rho\sigma} \partial_u( u \dC_{\rho\sigma})
		\nonumber\\
		&
		-
		N^{\rho\sigma} N_{\rho\sigma}
		(2\eth^{[\mu}C^- - C^- \bar{n}^{[\mu})n^{\nu]}
	\big],
\label{eq:loss_mech_J_rad_Lorentz}	
\end{align}
while its alternative form in Eq.~\eqref{eq:loss_mech_J_flux_inv} in terms of the invariant tensors $\iC_{AB}$ and $\iN_{AB}$ maps~onto
\begin{align}
	\Delta_{\J\rad}^{\mu\nu}
	=
	\int \frac{\dx u\dx^2\Omega}{32\pi G}
	\big[&
		4 \iC^{\rho[\mu}_{\mathstrut} \iN^{\nu]}_{\mathstrut}{}_\rho
		-
		2 \iN^{\rho\sigma} n^{[\mu}\eth^{\nu]} \iC_{\rho\sigma}
		\nonumber\\[-0.5em]&
		-
		n^{[\mu} \bar{n}^{\nu]}
		\iN^{\rho\sigma} \partial_u( u \iC_{\rho\sigma})
	\big].
\end{align}
Finally, the static term in Eq.~\eqref{eq:loss_mech_J_stat} becomes%
\begin{equation}
	\Delta_{\J\stat}^{\mu\nu}
	=
	\int\frac{\dx^2\Omega}{4\pi}
	M^+ (2\eth^{[\mu} \Delta S - \Delta S \bar{n}^{[\mu}) n^{\nu]}.
\label{eq:loss_mech_J_stat_Lorentz}	
\end{equation}

% ---------------------------------------------------------- %
\section[Boosted Schwarzschild metric]{Boosted\protect\\Schwarzschild metric}
\label{app:schwz}

In Sec.~\ref{sec:mech_schwz} of the main text, we motivated our definition of the mechanical angular momentum~$\J$ by appealing to the explicit form of the Bondi metric components $\{ M, N_A, C_{AB} \}$ for a boosted Schwarzschild spacetime. Those expressions, which are given in Eq.~\eqref{eq:schwz_metric_components}, are derived in this appendix. We~begin in Sec.~\ref{app:schwz_harmonic} by writing down the most general metric for a boosted Schwarzschild black hole in harmonic coordinates. The transformation to Bondi coordinates then proceeds in two stages. It is convenient to first transform the metric into Newman-Unti coordinates~\cite{Newman:1963ugj}, which we do in Sec.~\ref{app:schwz_NU}, before subsequently transforming to Bondi coordinates, which we do in Sec.~\ref{app:schwz_Bondi}. (This same set of transformations is discussed in Ref.~\cite{Blanchet:2020ngx} in the context of the multipolar post-Minkowskian expansion, which can be used to describe, e.g., the spacetime around an inspiraling binary.)

\subsection{Harmonic coordinates}
\label{app:schwz_harmonic}

Let ${\tilde{x}^\mu \equiv (\tilde{t}, \tilde{x}, \tilde{y}, \tilde{z})}$ denote a set of Lorentzian coordinates that satisfy the harmonic condition ${\tilde\partial_\mu(\sqrt{-\tilde{g}}\tilde g^{\mu\nu}) = 0}$, where $\tilde{g}^{\mu\nu}$ are the components of the inverse metric in this coordinate chart, $\tilde g$ is the determinant of its inverse, and ${\tilde\partial_\mu \equiv \partial/\partial \tilde{x}^\mu}$ is the partial derivative with respect to these coordinates. Analogously to how the Bondi coordinates $(u,r,\theta^A)$ have corresponding Lorentzian coordinates~${x^\mu \equiv (t,x,y,z)}$, we can introduce the retarded coordinates $(\tilde{u}, \tilde{r}, \tilde\theta^A)$ via
\begin{equation}
	(\tilde{t}, \tilde{x}, \tilde{y}, \tilde{z})
	=
	(\tilde{u}+\tilde{r},
	\tilde{r} \sin\tilde\theta\cos\tilde\phi,
	\tilde{r} \sin\tilde\theta\sin\tilde\phi,
	\tilde{r} \cos\tilde\theta).
\end{equation}
This correspondence allows us to define the null vector~$\tilde n^\mu$, the unit radial vector~$\hrv^\mu$, and the two basis vectors ${\tilde e^\mu_A}$, which are tangent to the unit 2-sphere, in the same way as how their Bondi counterparts $\{ n^\mu, \rv^\mu, e^\mu_A \}$ are defined in Appendix~\ref{app:Lorentz}, except that we here use the harmonic coordinates~$\tilde x^\mu$ in place of the Bondi coordinates~$x^\mu$. Just like their Bondi counterparts, these vectors are to be understood as living in the tangent bundle on~$\scri^+$, and so their indices are to be raised and lowered with the Minkowski metric~$\eta_{\mu\nu}$, not the~full~metric~$\tilde{g}_{\mu\nu}$.

Now consider the spacetime around a single Schwarzschild black hole of mass~$m$, whose center of energy travels along the worldline ${{\tilde x^\mu(\tau)} = b^\mu + p^\mu\tau/m}$. (We can set ${\eta_{\mu\nu} b^\mu p^\nu = 0}$ without loss of generality.) As~discussed in Sec.~\ref{sec:mech_schwz}, this worldline is only inferred via extrapolation, since the harmonic coordinates do not actually extend past the event horizon. More rigorously, the displacement vector~$b^\mu$ and the four-momentum~$p^\mu$~are defined as the constant vectors whose components parametrize the Poincar\'{e} transformation that takes us from the black hole's rest frame to this generic~inertial~frame.

By~starting with the inverse metric in the rest frame [see, e.g., Eq.~(5.172) of Ref.~\cite{doi:10.1017/CBO9781139507486}] and then performing this Poincar\'{e} transformation, we find that we can write
\begin{align}
	\tilde{g}^{\mu\nu}
	=&
	- \bigg( \frac{\brv + Gm}{\brv - Gm} \bigg) \frac{p^\mu p^\nu}{m^2}
	+
	\frac{\brv^2}{(\brv + Gm)^2} \Pi^{\mu\nu}
	\nonumber\\&
	-
	\frac{G^2 m^2}{(\brv + Gm)^2} \hat{\brv}^\mu \hat{\brv}^\nu,
\label{eq:schwz_harmonic_metric}
\end{align}
where the projection operator ${\Pi^{\mu\nu} = \eta^{\mu\nu} + p^\mu p^\nu/m^2}$, the scalar function ${\brv = \sqrt{ \Pi_{\mu\nu}(\tilde{x}^\mu - b^\mu)(\tilde{x}^\nu - b^\nu) }}$, and the spacelike vector ${\hat{\brv}^\mu = \Pi^\mu{}_\nu (\tilde{x}^\nu - b^\nu)/\brv}$; the indices on $\Pi^{\mu\nu}$ are lowered with the Minkowski metric. It is worth stressing that the expression in Eq.~\eqref{eq:schwz_harmonic_metric} is not generally covariant, as it holds only in harmonic coordinates; it is, however, Lorentz covariant. As a sanity check, note that if we undo the translation (i.e., send ${\tilde x^\mu \mapsto \tilde x^\mu + b^\mu}$) and then boost ourselves back into the black hole's rest frame such that ${p^\mu \to (m, \mathbf{0})}$, then ${\brv \to \tilde{r}}$ and ${\hat{\brv}^\mu \to \hrv^\mu}$, and we indeed recover Eq.~(5.172) of Ref.~\cite{doi:10.1017/CBO9781139507486}.

\subsection{Newman-Unti coordinates}
\label{app:schwz_NU}

As an intermediate step to obtaining the metric in Bondi coordinates, we shall first transform Eq.~\eqref{eq:schwz_harmonic_metric} into Newman-Unti coordinates~$(u, R, \theta^A)$.~These differ from the Bondi coordinates $(u,r,\theta^A)$ of the main text only by the choice of radial coordinate~\cite{Barnich:2011ty}; the retarded time~$u$ and the angular coordinates~$\theta^A$ are the same in both cases. To determine the relation between the harmonic coordinates $\tilde{x}^\mu$ and these Newman-Unti coordinates~$x^\id{a} \equiv (u,R,\theta^A)$, we use the standard transformation~law
\begin{equation}
	g^{\id{a}\id{b}}(x)
	=
	\frac{\partial x^\id{a}}{\partial \tilde{x}^\mu}
	\frac{\partial x^\id{b}}{\partial \tilde{x}^\nu}
	\tilde{g}^{\mu\nu}(\tilde x)
\label{eq:schwz_transformation_law}
\end{equation}
along with the gauge conditions
\begin{equation}
	g^{uu} = g^{uA} = 0,
	\quad
	g^{uR} = -1.
\label{eq:NU_gauge}
\end{equation}
Sans~serif indices $\{ \id{a}, \id{b}, \dots \}$ are used to emphasize that we are working in a nonrectangular coordinate chart. The fact that it is easier to impose the condition ${g^{uR} = -1}$ and then later transform from $R$ to~$r$, as opposed to imposing the Bondi gauge condition ${\partial_r \det(g_{AB}/r^2) = 0}$ directly, is why the Newman-Unti coordinates are a useful intermediate step. 

The metric in Eq.~\eqref{eq:schwz_harmonic_metric} is an exact solution to the vacuum Einstein equations, but to solve for $x^\id{a}$ as a function of~$\tilde{x}^\mu$, it is helpful to perform a post-Minkowskian expansion. We can do this because we are ultimately interested only in the metric components $\{ M, N_A, C_{AB} \}$, and so it suffices to determine just the first few terms in the $1/r$ expansion of the Bondi metric. Since a boosted Schwarzschild spacetime must admit smooth ${ u \to 0}$ and ${b^\mu \to 0}$ limits, it follows from dimensional analysis that Newton's constant~$G$ only ever appears in the Bondi metric as part of the dimensionless combination $G m/r$. This means that terms that are of higher order in $G$ are also of higher order in the~$1/r$ expansion. Because $M$ and $C_{AB}$ enter the Bondi metric starting at $O(1/r)$ [see~Eq.~\eqref{eq:Bondi_metric_expansion}], we only need to work up to $O(G)$ to determine these two quantities exactly. However, we will need to work up to $O(G^2)$ to determine $N_A$ exactly, since this quantity first enters the metric starting at~$O(1/r^2)$. We~therefore~write
\begin{equation}
	\tilde{g}^{\mu\nu} = \eta^{\mu\nu} - \sum_{n\geq 1} G^n \tilde{h}_n^{\mu\nu},
\end{equation}
with the first two terms in this expansion given by
\begin{subequations}
\label{eq:schwz_harmonic_metric_expanded}
\begin{align}
	\tilde{h}_1^{\mu\nu}
	&=
	\frac{2m}{\brv}
	\bigg( \eta^{\mu\nu} + \frac{2 p^\mu p^\nu}{m^2} \bigg),
	\label{eq:schwz_harmonic_metric_O1}
	\\
	\tilde{h}_2^{\mu\nu}
	&=
	\frac{m^2}{\brv^2}
	\bigg(
		\hat{\brv}^\mu\hat{\brv}^\nu
		-
		3 \eta^{\mu\nu}
		-
		\frac{p^\mu p^\nu}{m^2}
	\bigg).
\end{align}
\end{subequations}
The result for $\tilde{h}_1^{\mu\nu}$ can also be found in Refs.~\cite{Veneziano:2022zwh, Javadinezhad:2022hhl}.

We now assume that the relation between $x^\id{a}$ and $\tilde x^\mu$ can also be expanded perturbatively in powers of~$G$. Substituting the~ansatz
\begin{subequations}
\label{eq:harmonic_to_NU_ansatz}
\begin{align}
	u &= \tilde{u} + \sum_{n\geq 1} G^n \zeta^u_n(\tilde x),
	\label{eq:harmonic_to_NU_ansatz_u}
	\\
	R &= \tilde{r} + \sum_{n\geq 1} G^n \zeta^R_n(\tilde x),
	\label{eq:harmonic_to_NU_ansatz_R}
	\\
	\theta^A &= \tilde{\theta}^A + \sum_{n\geq 1} G^n \zeta^A_n(\tilde x)
	\label{eq:harmonic_to_NU_ansatz_A}
\end{align}
\end{subequations}
into Eq.~\eqref{eq:schwz_transformation_law} and then imposing the four Newman-Unti gauge conditions in Eq.~\eqref{eq:NU_gauge}, we obtain a set of partial differential equations for $\zeta_n^\id{a}$ at each order~$n$ in $G$. We have~that%
\begin{subequations}
\label{eq:NU_eqs_O1}
\begin{align}
	\frac{\partial}{\partial\tilde{r}} \zeta^u_1
	&=
	-\frac{1}{2} \tilde h_1^{\mu\nu} \tilde{n}_\mu \tilde{n}_\nu,
	\\
	\frac{\partial}{\partial\tilde{r}} \zeta^R_1
	&=
	\tilde h_1^{\mu\nu} \tilde n_\mu \hrv_\nu
	+
	\hrv^\mu \tilde\partial_\mu \zeta_1^u,
	\\
	\frac{\partial}{\partial\tilde{r}} \zeta^A_1
	&=
	\frac{1}{\tilde{r}} \tilde{e}^A_\mu
	\big(
		\tilde h_1^{\mu\nu} \tilde{n}_\nu
		+
		\tilde\partial^\mu \zeta_1^u
	\big)
\end{align}
\end{subequations}
at first order in~$G$, while at second order, we find that
\begin{subequations}
\label{eq:NU_eqs_O2}
\begin{align}
	\frac{\partial}{\partial\tilde{r}} \zeta^u_2
	=&
	-\frac{1}{2} \tilde h_2^{\mu\nu} \tilde{n}_\mu \tilde{n}_\nu
	+
	\frac{1}{2} \tilde\partial^\mu\zeta_1^u \tilde\partial_\mu \zeta_1^u
	\nonumber\\&
	+
	\tilde h_1^{\mu\nu} \tilde{n}_\mu
	\tilde\partial_\nu \zeta_1^u,
	\\
	\frac{\partial}{\partial\tilde{r}} \zeta^R_2
	=&\;
	\tilde h_2^{\mu\nu} \tilde n_\mu \hrv_\nu
	+
	\hrv^\mu \tilde\partial_\mu \zeta_2^u
	+
	\tilde\partial^\mu\zeta_1^u
	\tilde\partial_\mu \zeta_1^R
	\nonumber\\&
	+
	\tilde h_1^{\mu\nu} \tilde{n}_\nu
	\tilde\partial_\mu \zeta_1^R
	-
	\tilde{h}_1^{\mu\nu} \hrv_\mu \tilde\partial_\nu \zeta_1^u,
	\\
	\frac{\partial}{\partial\tilde{r}} \zeta^A_2
	=&\;
	\frac{1}{\tilde{r}} \tilde{e}^A_\mu
	\big(
		\tilde h_2^{\mu\nu} \tilde{n}_\nu
		+
		\tilde\partial^\mu \zeta_2^u
		-
		\tilde h_1^{\mu\nu} \tilde\partial_\nu\zeta_1^u
	\big)
	\nonumber\\&
	+
	\tilde\partial^\mu \zeta_1^u \tilde\partial_\mu \zeta_1^A
	+
	\tilde h_1^{\mu\nu} \tilde{n}_\mu \tilde\partial_\nu \zeta_1^A.
\end{align}
\end{subequations}
These two sets of equations can also be found in Ref.~\cite{Blanchet:2020ngx}.

Notice that each line in Eqs.~\eqref{eq:NU_eqs_O1} and~\eqref{eq:NU_eqs_O2} is a linear differential equation whose most general solution must therefore be the sum of a particular integral and a complementary function. These complementary functions, which are all independent of~$\tilde{r}$, account for two types of residual gauge freedoms: BMS~transformations, and the freedom to shift the origin of the radial coordinate~$R$~\cite{Barnich:2011ty, Blanchet:2020ngx}. The latter gauge freedom is present because the Newman-Unti radius~$R$ is the affine parameter along null geodesics with ${\dx u = \dx\theta^A = 0}$; the Bondi radius~$r$ exhibits no such residual freedom because it is not an affine parameter. Given that we eventually want to go into Bondi coordinates, we shall fix this residual gauge freedom in~$R$ by imposing the boundary condition given above Eq.~(2.4) of Ref.~\cite{Barnich:2011ty}, which is tantamount to requiring that ${\det(g_{AB}/R^2) = \det\Omega_{AB} + O(1/R^2)}$. For the remaining complementary functions associated with the BMS~group, we shall set them all to zero for the time being. The arbitrary supertranslation~$\beta$, which plays a key role in the discussion of Sec.~\ref{sec:mech}, can be added in at the end of the calculation by using the results in Eq.~(2.18) of~Ref.~\cite{Flanagan:2015pxa}.

The solution to Eq.~\eqref{eq:NU_eqs_O1} subject to these boundary conditions is~given~by (see also Refs.~\cite{Blanchet:2020ngx, Veneziano:2022zwh, Javadinezhad:2022ldc})
\begin{subequations}
\begin{align}
	\zeta_1^u &= 2(\tilde n \cdot p) \log \tilde r + O(\tilde r^{-1}),
	\\
	\zeta_1^R &= (-\tilde n \cdot p) + (\hrv \cdot p)(4-2\log\tilde r) + O(\tilde r^{-1}),
	\\
	\tilde{r} \zeta_1^A &= 2\tilde e^A_\mu p^\mu (1-\log\tilde r) + O(\tilde r^{-1}).
\end{align}
\end{subequations}
Only the leading terms in a $1/\tilde{r}$ expansion are presented for the sake of readability, but it is necessary to determine $\smash{\zeta_1^u}$, $\smash{\zeta_1^R}$, and $\smash{\tilde{r}\zeta_1^A}$ down to $\smash{O(\tilde{r}^{-2})}$ in order to correctly read off $\{ M, N_A, C_{AB} \}$. For the second-order equations in Eq.~\eqref{eq:NU_eqs_O2}, we must also determine $\zeta_2^u$, $\zeta_2^R$, and $\tilde{r}\zeta_2^A$ down to $O(\tilde{r}^{-2})$.

Having solved for~$\zeta_n^\id{a}$, we may now substitute Eq.~\eqref{eq:harmonic_to_NU_ansatz} back into Eq.~\eqref{eq:harmonic_to_NU_ansatz} to determine the inverse metric in Newman-Unti coordinates.~We note, of course, that because Eq.~\eqref{eq:harmonic_to_NU_ansatz} gives us $x^\id{a}$ as a function of~$\tilde{x}^\mu$, and because $\tilde{g}^{\mu\nu}$ is also a function of~$\tilde{x}^\mu$, this substitution returns the components of the Newman-Unti metric as functions of the harmonic coordinates~$\tilde{x}^\mu$. It~then remains to perform the inverse of the transformation in Eq.~\eqref{eq:harmonic_to_NU_ansatz} to express these components properly as functions of the Newman-Unti coordinates. Since $(u,R,\theta^A)$ are equal to $(\tilde u, \tilde r, \tilde\theta^A)$ at zeroth order in~$G$, we will need the inverse of Eq.~\eqref{eq:harmonic_to_NU_ansatz} only up to first order. Equations~\eqref{eq:harmonic_to_NU_ansatz_u} and \eqref{eq:harmonic_to_NU_ansatz_R} are easy enough to invert, and we get
\begin{subequations}
\begin{align}
	\tilde u &= u - G \zeta_1^u(x) + O(G^2),
	\\
	\tilde r &= R - G \zeta_1^R(x) + O(G^2).
\end{align}

For the angular coordinates, what we really need to know is how the basis vectors in the two coordinate systems are related, since all of the angular dependence in Eqs.~\eqref{eq:schwz_harmonic_metric}, \eqref{eq:NU_eqs_O1}, and \eqref{eq:NU_eqs_O2} arise from inner products like $(\tilde n \cdot p)$, $(\hrv \cdot b)$, and so~on. By first substituting Eq.~\eqref{eq:harmonic_to_NU_ansatz_A} into the definition for~$n^\mu$ in Eq.~\eqref{eq:n_def}, we find that $\smash{n^\mu = \tilde{n}^\mu + G \zeta_1^A(\tilde x) \partial\tilde{n}^\mu/\partial\tilde\theta^A + O(G^2)}$. Inverting this relation then tells us that
\begin{equation}
	\tilde n^\mu = n^\mu - G \zeta_1^A(x) e_A^\mu + O(G^2),
\end{equation}
after having used the definition of $e_A^\mu$ in Eq.~\eqref{eq:e_A}. Likewise,
\begin{equation}
	\hrv^\mu = \rv^\mu - G \zeta_1^A(x) e_A^\mu + O(G^2).
\end{equation}
In a similar way, we substitute Eq.~\eqref{eq:harmonic_to_NU_ansatz_A} into Eq.~\eqref{eq:e_A} and then use the chain rule to eventually find that
\begin{equation}
	\tilde e_\mu^A = e_\mu^A - G[\zeta_1^B(x) e_B^\nu \eth^{\vphantom{A}}_\nu e^A_\mu - \zeta_1^A(x) \rv_\mu] + O(G^2).
\end{equation}
\end{subequations}

\subsection{Bondi coordinates}
\label{app:schwz_Bondi}
Given the metric in Newman-Unti coordinates~$(u,R,\theta^A)$, we define the Bondi radius~$r$ via~\cite{Barnich:2011ty}
\begin{equation}
	r = \bigg(\frac{\det g_{AB}(u,R,\theta^C)}{\det \Omega_{AB}(\theta^C)}\bigg)^{1/4}.
\label{eq:NU_to_Bondi_r}
\end{equation}
This relation could now be used to transform our explicit result for a boosted Schwarzschild metric into Bondi coordinates, but in practice, it is easier to use Eq.~\eqref{eq:NU_to_Bondi_r} to transform the general Bondi metric in Eq.~\eqref{eq:Bondi_metric} into Newman-Unti coordinates. Having done so, one finds that for any nonradiative spacetime with ${N_{AB} = 0}$~\cite{Blanchet:2020ngx},
\begin{subequations}
\begin{align}
	g^{RR}
	=&\;
	1 - \frac{2GM}{R} + O(R^{-2}),
	\\
	R^2 g^{AB}
	=&\;
	\Omega^{AB} - \frac{1}{R}C^{AB} + O(R^{-2}),
	\\
	R g^{RA}
	=&\;
	\frac{1}{2R} D_B C^{AB}
	+
	\frac{2}{3R^2} G N^A
	\nonumber\\&
	-
	\frac{1}{2R^2} C^{AB} D^C C_{BC}
	+
	O(R^{-3}).
\end{align}
\end{subequations}
Comparing these general formulas with our explicit expressions for the boosted Schwarzschild metric allows us to read off the desired result for $\{M, N_A, C_{AB} \}$ as given in~Eq.~\eqref{eq:schwz_metric_components}.

% ---------------------------------------------------------- %
\section[Deriving the loss of mechanical angular momentum]%
{Deriving the loss of\protect\\mechanical angular momentum}
\label{app:proofs}

This appendix is divided into three parts.~We begin by listing a number of useful identities in Sec.~\ref{app:proofs_ids}.~The result in Eq.~\eqref{eq:j_deltaM_Cplus_final}, which is an integral step in the derivation of the radiated flux~$\smash{\Delta_\J^\rad}$, is then proved in~Sec.~\ref{app:proofs_flux}. Finally, in~Sec.~\ref{app:proofs_mem}, we prove the result in Eq.~\eqref{eq:loss_mech_J_mem_rewrite} for the~static~term~$\smash{\Delta_\J^\stat}$.

\subsection{Useful identities}
\label{app:proofs_ids}

To streamline our discussion, we begin by listing a number of identities that will be essential in later parts of this appendix. For starters, we have that
\begin{equation}
	(D_A D_B - D_B D_A) X_C
	=
	2\Omega_{A[C} \Omega_{D]B} X^D
\label{eq:commutator}
\end{equation}
for any vector $X^A$. This identity follows from the definition of the Riemann tensor and the fact that ${R_{ABCD} = 2\Omega_{A[C} \Omega_{D]B}}$ in the case of a round 2-sphere.

Next up are several identities for the conformal Killing vector~$Y^A$. First, we note that the contraction of Eq.~\eqref{eq:conformal_Killing} with any symmetric and traceless tensor~$X^{AB}$ yields
\begin{equation}
	X^{AB} D_A Y_B = 0 .
\label{eq:id_Y_1}
\end{equation}
Second, if we first differentiate Eq.~\eqref{eq:conformal_Killing}, we get
\begin{align}
	D_C D_B Y_A
	&=
	\Omega_{AB} D_C D_D Y^D - D_C D_A Y_B 
	\nonumber\\
	&=
	\Omega_{AB} D_C D_D Y^D - D_A D_C Y_B - 2 \Omega_{B[C}Y_{A]},
\end{align}
where the second line follows from using Eq.~\eqref{eq:commutator}. Now, symmetrizing over the indices $B$ and~$C$ and using Eq.~\eqref{eq:conformal_Killing}, we~obtain
\begin{align}
	D_{(B} D_{C)} Y_A
	=&\;
	\Omega_{A(B} D_{C)} D_D Y^D - \frac{1}{2} \Omega_{BC} D_A D_D Y^D
	\nonumber\\&
	- \Omega_{BC}Y_{A}
	+ \Omega_{A(B}Y_{C)}.
\end{align}
Contracting this with the symmetric and traceless tensor~$X^{BC}$ then~yields
\begin{equation}
	X^{BC} D_B D_C Y^A = X^{AB} D_B D_C Y^C + X^{AB} Y_B.
\label{eq:id_Y_2}
\end{equation}
A third identity involving three derivatives on~$Y^A$ reads~\cite{Elhashash:2021iev}
\begin{equation}
	X^{AB} D_A D_B D_C Y^C = 0.
\label{eq:id_Y_3}
\end{equation}

Finally, we will also make use of the fact that~\cite{Compere:2019gft}
\begin{equation}
	Y_{AB} D_C X^{BC} = 2 Y^{BC} D_{[A} X_{B]C}.
\label{eq:id_STF_pairs}
\end{equation}
for any pair of symmetric and traceless tensors, $X^{AB}$~and~$Y^{AB}$.

\subsection{Radiative term}
\label{app:proofs_flux}

Here we prove the result for $j(\Delta M, C^-)$ in Eq.~\eqref{eq:j_deltaM_Cplus_final}, which we use in the main text to obtain the radiated flux~$\smash{\Delta_\J^\rad}$. Our starting point is Eq.~\eqref{eq:j_deltaM_C-}, into which we substitute the expression for~$\Delta M$ in Eq.~\eqref{eq:DeltaM}. Having done so, we see that there are two types of terms in the result: those proportional to~$\Delta C_{AB}$, and those proportional to~$\Delta\mathcal{E}$. The former~read
\begin{equation}
	\int\frac{ \dx^2\Omega }{32\pi G}
	\Delta C^{AB} D_A D_B
	(2 Y^C D_C C^- - C^- D_C Y^C),
\label{eq:j_deltaM_C-_lin_1}
\end{equation}
where we have already integrated by parts twice to move the derivatives off~$\Delta C_{AB}$.~After using the product rule to distribute these derivatives, we find that several terms vanish or cancel one another due to the identities in Eqs.~\eqref{eq:id_Y_1}, \eqref{eq:id_Y_2}, and~\eqref{eq:id_Y_3}. For the terms that survive, we use Eq.~\eqref{eq:commutator} to show that
${
	D_A D_B D_C C^- = D_C D_A D_B C^- + 2 \Omega_{A[B} D_{C]}C^-
}$,
and thus find that Eq.~\eqref{eq:j_deltaM_C-_lin_1} is equivalent~to
\begin{align}
	\int \frac{ \dx^2\Omega }{32\pi G}
	\Delta C^{AB}
	\big( &
		2 Y^C D_C D_A D_B C^-
		- D_C Y^C D_A D_B C^-
		\nonumber\\[-0.5em]&
		+
		4 D_A Y^C D_B D_C C^-
	\big).
\label{eq:j_deltaM_C-_lin_2}	
\end{align}
Now use Eq.~\eqref{eq:twobody_CAB} to rewrite $D_A D_B C^-$ in terms of~$C^-_{AB}$. Integrating by parts and using $\smash{\Delta C_{AB} \equiv \int_{-\infty}^{\infty} N_{AB} \,\dx u}$ then~yields
\begin{align}
	-
	\int\frac{ \dx u \dx^2\Omega }{32\pi G} Y^A
	\bigg( &
		N^{BC} D_A C^-_{BC} - 2 D_B(N^{BC} C_{AC}^-)
		\nonumber\\&
		+
		\frac{1}{2} D_A (N^{BC} C^-_{BC})
	\bigg).
\label{eq:j_deltaM_C-_lin_3}
\end{align}

To complete the derivation, we must add to Eq.~\eqref{eq:j_deltaM_C-_lin_3} the terms in $j(\Delta M, C^-)$ that are proportional to~$\Delta\mathcal{E}$. They~read
\begin{align}
	&
	-\int\frac{\dx^2\Omega}{8\pi}
	(2 Y^A D_A C^- - C^- D_A Y^A)
	\Delta\mathcal{E}
	\nonumber\\
	=&
	-
	\int\frac{ \dx u \dx^2\Omega }{32\pi G}
	\bigg( Y^A D_A C^- - \frac{1}{2} C^- D_A Y^A \bigg)
	N^2
	\nonumber\\
	=&
	-
	\int\frac{ \dx u \dx^2\Omega }{32\pi G} Y^A
	\bigg(
		N^2 D_A C^-
		+
		\frac{1}{2} D_A (N^2  C^-)
 	\bigg).
\label{eq:j_deltaM_Cplus_quad}
\end{align}
The first equality follows from using the definition of $\Delta\mathcal{E}$ below Eq.~\eqref{eq:DeltaM} and writing ${N^2 \equiv N^{AB} N_{AB}}$ for brevity, while 
the second follows from integrating by parts. The sum of Eqs.~\eqref{eq:j_deltaM_C-_lin_3} and \eqref{eq:j_deltaM_Cplus_quad} gives us the desired result in Eq.~\eqref{eq:j_deltaM_Cplus_final}.

\subsection{Static term}
\label{app:proofs_mem}

Here we show that the two expressions for $\smash{\Delta_\J^\stat}$ in  Eqs.~\eqref{eq:loss_mech_J_stat} and~\eqref{eq:loss_mech_J_mem_rewrite} are equivalent. We start by inserting the expression for~$M^+$ in Eq.~\eqref{eq:def_M_via_f} into the former. After a straightforward integration by parts to move the derivatives off~$f^+_{AB}$, we find that $\smash{\Delta_\J^\stat}$ is equal~to
\begin{align}
	\int & \frac{\dx^2\Omega}{8\pi}
	\bigg(
		\frac{1}{4G} f^+_{AB} D^A D^B
		(2 Y^C D_C \Delta S - \Delta S D_C Y^C)
		\nonumber\\&
		+
		(3\rv - \hat t)\cdot P^+
		(2 Y^C D_C \Delta S - \Delta S D_C Y^C)
	\bigg).
\label{eq:flux_mech_J_soft_temp_1}
\end{align}
Now notice that the first line above is structurally identical to Eq.~\eqref{eq:j_deltaM_C-_lin_1}, and so all of the steps taken in the previous part of this appendix follow through. The only key difference is that~$\Delta Z$ lives in the kernel of the differential operator $(2D_A D_B - \Omega_{AB}D^2)$, and so drops out when we integrate by parts. Ultimately, we arrive at the expression
\begin{align}
	\int & \frac{\dx u \dx^2\Omega}{32\pi G}
	Y^A
	\bigg(
		{-} f_+^{BC} D_A N_{BC} + 2 D_B(f_+^{BC} N_{AC})
		\nonumber\\&
		\qquad\qquad\;\;
		-
		\frac{1}{2} D_A (f_+^{BC} N_{BC})
	\bigg)
	\nonumber\\
	+
	\int & \frac{\dx^2\Omega}{8\pi}
	\bigg(
		3 (\rv \cdot P^+)
		(2 Y^C D_C \Delta S - \Delta S D_C Y^C)
		\nonumber\\&\quad\quad
		+
		3 (\hat t\cdot P^+) (D_A Y^A) \Delta S 
	\bigg).
\label{eq:flux_mech_J_soft_temp_2}
\end{align}
To get the fourth line, we have also performed one more integration by parts while using the fact that ${(\hat t \cdot P^+)}$ is a constant. Equation~\eqref{eq:flux_mech_J_soft_temp_2} is still not quite the desired result, however. Observe that we essentially have to exchange $\smash{ (f^+_{AB}, N^{\mathstrut}_{AB}) \mapsto (- N^{\mathstrut}_{AB}, f^+_{AB}) }$ to go from this to Eq.~\eqref{eq:loss_mech_J_mem_rewrite}. We do so by using Eq.~\eqref{eq:id_STF_pairs} to show that
\begin{equation}
	D_B(f^{BC}_+ N_{AC}) = D_A(N^{BC} f^+_{BC}) - D_B(N^{BC} f^+_{AC}).
\end{equation}
Inserting this into~Eq.~\eqref{eq:flux_mech_J_soft_temp_2} achieves the desired~outcome.

% ---------------------------------------------------------- %
\section[Other definitions of angular momentum]%
{Other definitions\protect\\of angular momentum}
\label{app:comparisons}

% Spacing for TOC in arXiv version
\addtocontents{toc}{~}

Several references~\cite{Compere:2019gft, Chen:2021szm, Chen:2021kug, Mao:2023evc} have proposed a super\-translation-invariant definition of the angular momentum that, in our notation, reads
\begin{equation}
	J_\text{(inv)}(\sigma)
	\coloneq
	J(\sigma) - j(M(\sigma), C(\sigma)).
\label{eq:J_CWWY}
\end{equation}
Both arguments of the function $j(\cdot,\cdot)$, which is as defined in Eq.~\eqref{eq:def_little_j}, are to be evaluated on the cut~$\sigma$. In the general case, the shear tensor need not have the form in Eq.~\eqref{eq:schwz_metric_components_CAB}, but the scalar potential~${C \equiv C(u,\theta^A)}$ can still be extracted from $C_{AB}$ via the more general decomposition
\begin{equation}
	C_{AB}
	=
	-(2 D_A D_B - \Omega_{AB} D^2) C
	+
	\epsilon_{C(A} D_{B)} D^C \bar C,
\end{equation}
where $\epsilon_{AB}$ is the volume form on~$\sigma$ and ${ \bar C \equiv \bar C(u,\theta^A) }$ is a second scalar potential, sometimes called the magnetic-parity piece of the shear~\cite{Flanagan:2015pxa}.~(In this terminology, $C$ is the electric-parity piece of the shear.)

From Eqs.~\eqref{eq:def_mech_J_binary} and~\eqref{eq:twobody_Z_initial}, we see that $J_\text{(inv)}$ coincides with our definition for the mechanical angular momentum~$\J$ on the initial cut~$\scri^+_-$, but $J_\text{(inv)}$ and $\J$ are inequivalent on the final cut~$\scri^+_+$ because of the shift in reference point~$\Delta Z$ that we have corrected for. Consequently, If we let $\smash{\Delta_\text{(inv)}\coloneq J^-_\text{(inv)} - J^+_\text{(inv)}}$ denote the total loss of $J_\text{(inv)}$ between $\scri^+_-$ and~$\scri^+_+$, then the difference between $\Delta_\text{(inv)}$ and our result for~$\Delta_\J$~is
\begin{equation}
	\Delta_\J - \Delta_\text{(inv)}
	=
	j(M^+, \Delta Z).
\label{eq:comparison_loss_inv}
\end{equation}
Now appropriating the results of Appendix~\ref{app:proofs}, we find that Eq.~\eqref{eq:comparison_loss_inv} is equivalent to
\begin{align}
	\int\frac{\dx^2\Omega}{8\pi} &
	\bigg(
		3 (\rv \cdot P^+)
		(2 Y^C D_C \Delta Z - \Delta Z D_C Y^C)
		\nonumber\\&
		+
		3 (\hat t\cdot P^+) (D_A Y^A) \Delta Z
	\bigg).
\end{align}
Reasoning similar to that below Eq.~\eqref{eq:loss_mech_J_mem_rewrite} of Sec.~\ref{sec:pm_cm} tells us that the space-space components of $\Delta_\text{(inv)}$ and $\Delta_\J$ agree at $O(G^2)$ in the binary's \cm~frame, but otherwise these two quantities are generally inequivalent. Moreover, as discussed in Sec.~\ref{sec:pm_stat}, the quantity $j(M^+, \Delta Z)$ is not Lorentz covariant because projection operators like $\mathbb{P}_{\ell\leq 1}$ and $\mathbb{P}_{\ell\geq 2}$ applied to $\Delta S$ do not commute with Lorentz boosts. Since we can verify explicitly that $\Delta_\J$ is Lorentz covariant, Eq.~\eqref{eq:comparison_loss_inv} implies that $\Delta_\text{(inv)}$ does not transform covariantly under boosts.

A different proposal for a supertranslation-invariant definition of the angular momentum was put forward by Javadinezhad and Porrati~(JP) in Ref.~\cite{Javadinezhad:2022ldc}. Their definition~reads
\begin{equation}
	J_\text{(JP)}(\sigma)
	=
	J(\sigma) - j(M(\sigma), C^-) + j(M(\sigma), \mathcal{C}). \bigg.
\label{eq:def_J_JP}
\end{equation}
[Actually, the definition in Ref.~\cite{Javadinezhad:2022ldc} is valid only for ${D_A Y^A = 0}$; in the Lorentz-tensor representation, this would correspond to defining just the space-space components~$\smash{ J_\text{(JP)}^{ij} }$. To facilitate a clearer comparison with our own result, we have offered a natural extension of their definition that holds in the more general case ${D_A Y^A \neq 0}$.]

The effect of the second and third terms in Eq.~\eqref{eq:def_J_JP} is to remove the contribution of the initial shear~$C^-$ from the Bondi angular momentum~$J$ and replace it by the function~$\smash{\mathcal{C} \equiv \mathcal{C}(\theta^A)}$, which is chosen in such a way that the space-space components of ${\Delta_\text{(JP)} \coloneq J^-_\text{(JP)} - J^+_\text{(JP)}}$ yield the same result as the Bondi flux~$\smash{ F_J^{ij} }$ when the latter is computed in the intrinsic~gauge. It then follows from the discussion below Eq.~\eqref{eq:loss_mech_J_mem_rewrite} that $\smash{ \Delta_\text{(JP)}^{ij} }$ also agrees with the result of $\smash{ \Delta_\J^{ij} }$ at $O(G^2)$ in the intrinsic \cm~frame, but generally $\Delta_\text{(JP)}$ and $\Delta_\J$ are inequivalent. 
\bibliography{main}
\end{document}